\documentclass[11pt,titlepage]{article}
\usepackage{amssymb}
\usepackage{amsmath}
\usepackage{amsfonts}
\usepackage[doublespacing]{setspace}
\usepackage{geometry}

\evensidemargin=\oddsidemargin
\newdimen\dummy
\dummy=\oddsidemargin
\newtheorem{theorem}{Theorem}

\newtheorem{corollary}{Corollary}

\newtheorem{definition}{Definition}
\newtheorem{example}{Example}

\newtheorem{lemma}{Lemma}

\newenvironment{proof}[1][Proof]{\noindent\textbf{#1.} }{\ \rule{0.5em}{0.5em}}
\input{tcilatex}
\begin{document}

\title{Market Making with Model Uncertainty\thanks{%
We thank the participants of the seminar held in 2015 at the Department of
Management Science and Engineering, Stanford University for helpful
comments. }}
\author{Hee Su Roh\thanks{%
Research Scientist, Korea Advanced Institute of Science and Technology.
Email: roh6828@kaist.ac.kr}, Yinyu Ye\thanks{%
K.T.Li Chair Professor of Engineering, Department of Management Science and
Engineering, Stanford University. Email: yinyu-ye@stanford.edu}}
\date{November 10, 2015}
\maketitle

\begin{abstract}
Pari-mutuel markets are trading platforms through which the common market
maker simultaneously clears multiple contingent claims markets. This market
has several distinctive properties that began attracting the attention of
the financial industry in the 2000s. For example, the platform aggregates
liquidity from the individual contingent claims market into the common pool
while shielding the market maker from potential financial loss. The
contribution of this paper is two-fold. First, we provide a new economic
interpretation of the market-clearing strategy of a pari-mutuel market that
is well known in the literature. The pari-mutuel auctioneer is shown to be
equivalent to the market maker with extreme ambiguity aversion for the
future contingent event. Second, based on this theoretical understanding, we
present a new market-clearing algorithm called the Knightian Pari-mutuel
Mechanism (KPM). The KPM retains many interesting properties of pari-mutuel
markets while explicitly controlling for the market maker's ambiguity
aversion. In addition, the KPM is computationally efficient in that it is
solvable in polynomial time.
\end{abstract}

\section{Introduction}

In this paper, we design a new platform for trading contingent claims, the
Knightian Pari-mutuel Mechanism.

The term "pari-mutuel" originates from the automated horse race betting
system invented in the 19th century. The pari-mutuel betting system
automatically calculates payoff odds for each horse based on the amount of
money bet on each horse. It also completely shields the market organizer
from financial loss. To illustrate, suppose that people wager their money on
the outcome of a race between two horses: A and B. People wager a total of
\$50 on Horse A and \$100 on Horse B. The total premium of \$150 is paid to
those who correctly predicted the outcome. Thus, if Horse A wins, those who
wagered money on Horse A receive \$3 for each dollar they wagered. If Horse
B wins, the winners make \$1.5 for each dollar they wagered. Because the
payment to the winners is financed exclusively by the fees collected from
both winners and losers, the market maker does not need to worry about
his/her loss. This is called the self-financing property of the market.

The rate of return from wagering on a particular horse conveys information
on the collective perception of that horse's chances of winning. Consider
the example above. Wagering on Horse B yields a lower rate of return than
wagering on Horse A because people wagered more money on Horse B than on
Horse A. The more money people wager on a particular horse the lower the
rate of return becomes. People bet money on a horse if they believe that
horse is likely to win the race. Thus, the rate of return from wagering
money on a horse is low if many people believe that the horse will win the
race. The pari-mutuel system maps the popularity of horses to the rates of
return from wagering money on those horses.

This simple pari-mutuel system subsequently evolved into more sophisticated
prediction markets. For example, more recently developed markets (Peters et
al., 2005; Peters et al., 2007) trade securities with fixed final payoffs.
The prices of those securities fluctuate in a way that reflects their
popularity in the market.

Despite considerable heterogeneity across various prediction markets, they
typically exhibit three defining characteristics. First, the popularity of a
particular security is mapped to a higher price of that security through an
automated market-clearing algorithm. Second, the market maker's maximum
possible loss at maturity is bounded. Irrespective of the outcome at the
time when contingent claims mature, the market maker is not expected to lose
more than a certain pre-specified amount. In this paper, when we say that
the market is \textit{completely} pari-mutuel, we mean that the market maker
is not expected to lose money, regardless of the outcome.

Finally, the market aggregates liquidity across different markets into the
common pool (Baron and Lange, 2007). For example, consider the horse race
example above but with a slight modification. Suppose that people trade
securities with fixed payoffs: Claim A and Claim B. Let "Claim A" refer to
the contingent claim that pays \$1 if and only if Horse A wins. Define the
term "Claim B" similarly. The potential payout to the holders of Claim A is
financed by the premiums collected from the holders of Claim B and vice
versa. Therefore, it is as if the holders of Claim A and those of Claim B
were transacting with one another via a common pari-mutuel auctioneer.
Compare this case with an alternative situation in which potential buyers of
Claim A (or Claim B) only trade with potential sellers of Claim A (or Claim
B). The effective number of people trading with one another is larger in the
former case. It is as if the common auctioneer pooled liquidity from
individual markets - the market for Claim A and the market for Claim B -
into the common pool. As a result, market participants can enjoy a more
liquid market.

In the 2000s, researchers (Lange and Economide, 2005; Baron and Lange, 2007)
noted that the pari-mutuel principle can be used to better organize a
certain type of financial derivatives market. For example, note that the
pari-mutuel auctioneer is well protected from financial loss at the time the
claims mature. This property of the pari-mutuel market allows the auctioneer
to be less concerned with fluctuations in the value of the inventory.
Therefore, the pari-mutuel principle can be used to design a market if the
market maker has difficulty hedging against inventory risk (Lange and
Economide, 2005; Baron and Lange, 2007). For example, Lange and Economide
(2005) designed the Pari-mutuel Digital Call Auction (PDCA) to trade options
written on economic indices, for which delta hedging using the underlying
asset is not feasible.

Longitude, a financial technology company, developed software to implement
the PDCA. In collaboration with investment banks (e.g., Goldman Sachs) and
financial exchanges (e.g., the International Securities Exchange (ISE)) new
PDCA-based derivatives markets were launched. Due to lack of active market
participation, the ISE shut down the auction in June 2007. However, the ISE
has shown consistent interest in utilizing this technology in the near
future (Burne, 2013).

Despite their use in the financial industry, many pari-mutuel auctions have
design features that are different from the modeling assumptions that
economists use. A potential reason for this is that pari-mutuel auctions
have primarily been studied by scholars in operations research. For example,
many pari-mutuel auctions optimally clear the market while placing a lower
bound on the auctioneer's maximum possible loss (Hanson, 2003; Pennock,
2004; Peters et al., 2005; Lange and Economide, 2005; Chen and Pennock,
2007; Peters et al., 2007; Abernethy et al., 2013). The worst-case scenario
can have a material impact on how the auctioneer clears the market even if
such a scenario is very unlikely. In contrast, the market maker in the
economists' model is often exclusively concerned with maximizing his/her 
\textit{expected }utility derived from the monetary payoff. When only the 
\textit{expected value }of the future utility is concerned, extreme
worst-case loss with a small probability of occurrence does not merit
considerable attention.

The contribution of our paper is two-fold. First, we present a theoretical
framework through which pari-mutuel auctions can be reconciled with standard
economic models. Regarding the economics model, we focus on a market maker
with extreme ambiguity aversion for the future contingent event on which
claims are written. The decision maker with ambiguity aversion is uncertain
of which probability distribution accurately describes the contingent event.
For a pari-mutuel market, we\ consider the Convex Pari-mutuel Call Auction
Mechanism (Peters et al., 2005), which is an improved version of the PDCA.
We show that the market-clearing strategy of the market maker with extreme
ambiguity aversion is asymptotically equivalent to that of the CPCAM
auctioneer. By \textit{asymptotic equivalence, }we imply making the CPCAM
increasingly completely pari-mutuel.\footnote{%
In the CPCAM, before the beginning of the regular trading session, the
initial liquidity provider seeds the market with small initial orders. This
initial order, which is typically called the starting order, is a unique
design feature of the CPCAM. The starting order is introduced into the CPCAM
only to ensure the existence of unique state prices, which are used to
compute the market-clearing prices of contingent claims. However, the
starting order exposes the market organizer to a financial loss at the time
the claims mature. The larger the starting order, the greater the potential
financial loss of the market maker. By \textit{asymptotic equivalence, }I
mean reducing the magnitude of the starting order toward zero. When the
starting orders are infinitely small, the market-clearing strategy of the
CPCAM auctioneer approaches that of the ambiguity-averse market maker.
\par
In addition, I assume that the market organizer submits the same starting
order for all possible states of the future. I describe what the starting
order is in later sections.}

Second, based on this unified theoretical framework, we design a new market
called the Knightian Pari-mutuel Mechanism (KPM). The KPM has a solid
microeconomic rationale behind its design. We derive the optimization
problem of the KPM by modeling the market maker using the theory of decision
making under ambiguity aversion. The market-clearing algorithm explicitly
controls for the level of the market maker's ambiguity aversion. In
addition, we propose an algorithm that can compute an optimal solution to
the optimization problem in polynomial time.

\subsection{Literature Review}

In the prediction market literature, the KPM is most similar to Chen and
Pennock's utility-based market maker (Chen and Pennock, 2007). The
utility-based market maker prices contingent claims in such a way that the
transaction leaves the market maker's expected utility over the future
monetary payoff unaffected. Agrawal et al. (2011) suggests an improvement of
Chen and Pennock's (2007) market. The market maker may find it difficult to
propose a unique probability distribution to describe the event for which
the claims are written. The KPM addresses Agrawal et al. (2011)'s
suggestion: It\ allows the market maker to indicate a set of multiple
reasonable probability distributions instead of a single distribution. The
KPM acknowledges that the market maker often cannot pin down a single
subjective probability distribution.

Our paper is related to the recent literature that explains a wide variety
of prediction markets from a unified theoretical perspective. The most
notable work in this respect is Agrawal et al. (2011). They show that the
four most well-known prediction markets in the literature can be unified
under one theoretical framework. In a similar vein, we reconcile pari-mutuel
mechanisms from the prediction market literature with the model from the
economics literature.

My paper is related to a growing body of literature that focuses on the role
of Knightian uncertainty in decision making. In the past decade, Knightian
uncertainty has received a significant amount of attention in areas ranging
from macroeconomic modeling (Hansen and Sargent, 2008) to market
microstructure theory (Easley and O'Hara, 2009; Easley and O'Hara, 2010).

\section{The Theory of Decision Making under Ambiguity}

We present a brief overview of the theory of decision making under
ambiguity. First, it is necessary to distinguish between \textit{risk} and 
\textit{ambiguity. }Risk applies to situations in which it is possible to
attach a probability distribution to an unknown prospect. By contrast,
ambiguity refers to situations in which it is impossible to do so. For
example, consider a situation in which a person receives a dollar if and
only if he/she draws a red ball from a box. The box contains both red balls
and blue balls. If the person knows the fraction of balls that are red,
he/she knows the probability of receiving a dollar. In this case, the person
is said to be facing \textit{risk. }On the other hand, suppose that the
person does not know the fraction of balls in the box that are red. Then,
the person cannot assign a number to the probability of winning a dollar.
This person is said to be facing \textit{ambiguity. }In the 1920s, Knight
was the first to note the difference between these two concepts (Knight,
1936).

An \textit{ambiguous }prospect requires a different analysis from that of a
risky prospect. To this end, Theorem 1 reproduces the main finding of Gilboa
and Schmeidler (1989) in the language of Ghirardato et al. (2004). Let $S$
denote the set of all possible states (e.g., the person chooses a red ball,
the person chooses a blue ball), and let $X$ denote the set of consequences
(e.g., the person wins a dollar). Subsets of $S$ are called events. Let $%
\Sigma $ denote the algebra of subsets of the state space $X$. We are
interested in the decision maker's preference over different \textit{simple
acts}: A simple act is a $\Sigma $-measurable function $f:S\longrightarrow X$
that is finite-valued (Ghirardato e al, 2004). Let $\tciFourier $ denote the
set of all simple acts. Suppose that the binary relations $\succcurlyeq $
and $\succ $ characterize the decision maker's preference over different
acts: $f\succcurlyeq (\succ )g$ if and only if the decision maker (strictly)
prefers the simple act $f$ to the simple act $g$. Finally, let $%
u:X\rightarrow 
\mathbb{R}
$ denote the decision maker's utility function.

\begin{theorem}[Decision Making under Ambiguity]
\textbf{(Gilboa and Schmeidler, 1989; Ghirardato et al., 2004) }The decision
maker's preference relation $\succcurlyeq $ satisfies the set of six
behavioral axioms\footnote{%
Please see Ghirardato et al. (2004) for the set of six behavioral axioms.}
if and only if there exists a unique set $\Psi $ of probabilities on $%
(S,\Sigma )$ such that (\ref{GhirardatoEq}) holds for $\forall f,g\in
\tciFourier $. The set $\Psi $ is weakly compact, convex and nonempty.%
\footnote{%
Ghirardato et al. (2004) presents three different versions of the theorem
depending on the DM's attitude toward ambiguity. However, we only work with
the version that assumes \textit{aversion }to ambiguity. Please refer to
Ghirardato et al. (2004) for a more rigorous formal definition of aversion
to ambiguity.}%
\begin{equation}
f\succcurlyeq g\Leftrightarrow \min\limits_{P\in \Psi }\int u(f)dP\geq
\min\limits_{P\in \Psi }\int u(g)dP  \label{GhirardatoEq}
\end{equation}

\begin{proof}
See Gilboa and Schmeidler (1989) or Ghirardato et al. (2004)\textbf{.}
\end{proof}
\end{theorem}

The set of probabilities $\Psi $ in Theorem 1 encapsulates the decision
maker's (hereafter called the DM) perception of ambiguity (Ghirardato et
al., 2004). Recall that a DM facing ambiguity cannot attach a single
probability distribution to the unknown prospect. Instead, the DM has a set
of candidates $\Psi $ that he/she believes are fairly accurate predictions
of the future (Ghirardato et al., 2004). In other words, the DM has a set of
multiple priors (Gilboa and Schmeidler, 1989). The size of $\Psi $
represents the extent to which the DM feels ambiguous toward the unrealized
future outcome (Ghirardato et al., 2004). A large size of $\Psi $ implies
that the DM cannot easily narrow down the set of reasonable probability
distributions because he/she is too ambiguous about the future outcome
(Ghirardato et al., 2004).

Among the set of multiple priors, the DM is exclusively concerned with the
worst possible scenario. First, for each $P\in \Psi $, the DM calculates the
expected utility $\int u(f)dP$ from the unknown prospect assuming that $P$
is the true description of the future. Second, the DM finds the distribution
that results in the lowest level of utility. Third, when comparing one act
with another, the DM uses the probability distribution associated with the
worst scenario. The DM chooses the act whose worst-case scenario is better
than the worst-case scenarios of the other acts. See the Appendix for a
numerical example.

In practical modeling and implementation, the specification of the set $\Psi 
$ of the DM becomes another issue. Hansen and Sargent (2008) presents a
useful solution in the context of modeling in macroeconomics. We first
introduce Kullback's cross-entropy function (Cover and Thomas, 2012) to
quantify the extent to which two probability distributions differ from one
another.

\begin{definition}
\textbf{(Kullback's Cross-Entropy Function) }Suppose that there are two
probability distributions $p$ and $q$ with the common support set $S.$
Suppose that $q$ is the prior density over the set $S$. Then, the Kullback's
cross-entropy function is defined as (\ref{KullbackCrossEntropy}) (Cover and
Thomas, 2012).\ A large value of $S(p,q)$ implies that $p$ and $q$ are very
different from one another.%
\begin{equation}
S(p,q)=\int_{S}p(x)\ln \left[ \frac{p(x)}{q(x)}\right] dx
\label{KullbackCrossEntropy}
\end{equation}
\end{definition}

Hansen and Sargent (2008) use cross-entropy to restrict the set of
probability distributions considered by the DM. Given the prior distribution 
$q$ and a parameter $\eta $, the DM's set $\Pi $ includes all probability
distributions $p$ for which $S(p,q)\leq \eta $. As long as the probability
distributions are not too different from $p$, in which case $S(p,q)>\eta $,
the DM considers those probability distributions to be equally acceptable.

A large value of the parameter $\eta $ quantifies the DM's strong ambiguity
aversion.\footnote{%
Illeditsch (2011) also uses the size of the set of possible models under the
DM's consideration as a proxy for the DM's level of ambiguity aversion.}
With a larger value of $\eta $, the DM regards a larger set of probability
distributions as candidates for accurate descriptions of the world. Hence, a
large $\eta $ is equivalent to saying that the DM is more ambiguous about
the real world.

\section{The Microeconomic Analysis of the Convex Pari-mutuel Call Auction
Mechanism (CPCAM)}

\subsection{The Market Setting}

The CPCAM\ allows the common market maker to simultaneously handle different
types of contingent claims as long as the claims are written on the same
uncertain event (e.g., the outcome of the world cup, stock prices). Suppose
that there are $N$ possible outcomes of the uncertain event, each of which
is indexed by $i\in \{1,2,...,N\}$.

The CPCAM is a call auction. For simplicity, only buy orders are accepted.
Suppose that the market participants as a whole submit $J$ orders to the
market maker. Let the matrix $\mathbf{A\in 
\mathbb{R}
}^{N\times J}$ denote the payoff structure of $J$ orders. The $(i,j)$
element of $\mathbf{A}$ denotes the per-share payoff of the $j$th order,
where $j\in \{1,2,...,J\}$ if the $i$th outcome is realized. Define the
vector $\mathbf{b\in 
\mathbb{R}
}^{J}$ such that the $j$th element of this vector is the limit price
associated with the $j$th order. Define the vector $\mathbf{Q\in 
\mathbb{R}
}^{J}$ such that its $j$th element is the limit quantity for the $j$th order.

$\mathbf{\delta \in 
\mathbb{R}
}^{N}$ denotes the starting order. The starting order is a unique feature of
pari-mutuel auctions (Lange and Economide, 2005; Peters et al., 2005).
Before regular traders submit their orders, the market organizer seeds the
market with the starting order $\mathbf{\delta }$. For each $i$, the
organizer purchases $\delta _{i}$ dollars' worth of the Arrow-Debreu
security that pays \$1 per share if and only if the $i$th outcome is
realized. Arrow-Debreu securities are introduced only for the starting order
and thus are not traded in the regular trading session. Let "the $i$th
Arrow-Debreu security" refer to the one that pays \$1 per share if and only
if the $i$th event is realized. At this point, the organizer does not know
the number of shares of Arrow-Debreu securities he/she owns because those
securities are not yet priced. The prices of Arrow-Debreu securities are
determined only when the markets are cleared at the end of the regular
trading session. The number of the $i$th Arrow-Debreu security the organizer
holds is determined by dividing $\delta _{i}$ by the price of that security.
Then, the auctioneer pays the market organizer just like any other trader.
The starting orders are included in the model to ensure that the market
clearing optimization problem yields a unique set of prices for contingent
claims (Lange and Economide, 2005; Peters et al., 2005).

Equation (\ref{CPCAM}) is the CPCAM. Let $\mathbf{\varepsilon \in 
\mathbb{R}
}^{N}$ denote the vector of state prices: The $i$th element of $\mathbf{%
\varepsilon }$ is the state price for the $i$th outcome. $\varepsilon _{i}$
is the Lagrange multiplier associated with the constraint $%
\sum_{j=1}^{J}A_{i,j}x_{j}+s_{i}=M$. The state prices are the building
blocks on the basis of which all contingent claims traded on this market are
priced. For example, the market-clearing price for the contingent claim with
the payoff structure $\mathbf{A}_{\cdot j}$ is $\mathbf{A}_{\cdot j}^{T}%
\mathbf{\varepsilon }$. $\mathbf{s\mathbf{\in 
\mathbb{R}
}^{N}}$ and $M$ are dummy variables. $\mathbf{x\in 
\mathbb{R}
}^{J}$ is the vector of order fills. For example, the $j$th element of $%
\mathbf{x}$ is the number of shares of the claim that the submitter of the $%
j $th order is allowed to purchase.

\begin{equation}
\begin{array}{c}
\max\limits_{\mathbf{x,s,}M}\mathbf{b}^{T}\mathbf{x-}M+\sum_{i=1}^{N}\delta
_{i}\log (s_{i}) \\ 
\text{such that}\smallskip \\ 
\begin{tabular}{ll}
(A) & $\sum_{j=1}^{J}A_{i,j}x_{j}+s_{i}=M$ for each $i\in \{1,2,...,N\}$ \\ 
(B) & $\mathbf{0\leq x\leq Q}\smallskip $ \\ 
(C) & $\mathbf{s}\geq \mathbf{0\smallskip }$%
\end{tabular}%
\end{array}
\label{CPCAM}
\end{equation}

The Karush-Kuhn-Tucker (KKT) optimality condition for (\ref{CPCAM}) implies
the limit order logic (\ref{limitOrderLogic}) for each $j$. The market maker
can exercise his/her discretion if the bid price is exactly equal to the
market-clearing price of the order. 
\begin{equation}
\begin{array}{ccc}
x_{j}=0 & if & \mathbf{A}_{\cdot j}^{T}\mathbf{\varepsilon }>b_{j} \\ 
x_{j}\in \left[ 0,Q_{j}\right] & if & \mathbf{A}_{\cdot j}^{T}\mathbf{%
\varepsilon }=b_{j} \\ 
x_{j}=Q_{j} & if & \mathbf{A}_{\cdot j}^{T}\mathbf{\varepsilon }<b_{j}%
\end{array}
\label{limitOrderLogic}
\end{equation}

The person who submitted the $j$th order pays the premium worth $b_{j}x_{j}$
to the market maker. If the $i$th outcome is realized, the market maker pays
the person $A_{ij}x_{j}$.

The $\sum_{i=1}^{N}\delta _{i}\log (s_{i})$ term ensures the existence of a
unique state price vector. However, the starting order subjects the market
organizer to potential financial loss when the claims mature. To minimize
organizer's potential loss, Peters et al. (2005) suggest making the
magnitude of $\mathbf{\delta }$ very small.

\subsection{Equivalence with the Ambiguity-Averse Market Maker}

Let $u:%
\mathbb{R}
\rightarrow 
\mathbb{R}
$ denote the market maker's utility function. Suppose that $u$ is an
increasing function. Unlike Peters et al. (2005), we suppose that the market
maker uses the uniform starting order. That is, $\delta _{i}$ is the same
constant $\delta $ for $\forall i$. Let $\mathbf{\varepsilon }(\delta )$
denote the state price vector associated with (\ref{CPCAM}) when $\delta
_{i}=\delta $ for $\forall i$. Let $\mathbf{x}(\delta )$ denote an optimal
value of $\mathbf{x}$ for (\ref{CPCAM}).

\begin{theorem}
As $\delta \rightarrow 0$, $\mathbf{x}(\delta )$ converges to an optimal
solution for (\ref{AmbiguityAverseMarketMaker}). 
\begin{equation}
\begin{array}{c}
\max\limits_{\mathbf{x}}\min\limits_{\mathbf{p}}\sum\limits_{i=1}^{N}p_{i}u%
\left[ \mathbf{b}^{T}\mathbf{x-}\sum_{j=1}^{J}A_{i,j}x_{j}\right] \\ 
\text{such that}\smallskip \\ 
\begin{tabular}{ll}
(A') & $\mathbf{0\leq x\leq Q}\smallskip $ \\ 
(B') & $\mathbf{p\geq 0},\sum_{i=1}^{N}p_{i}=1$%
\end{tabular}%
\end{array}
\label{AmbiguityAverseMarketMaker}
\end{equation}

\begin{proof}
See the Appendix.
\end{proof}
\end{theorem}

(\ref{AmbiguityAverseMarketMaker}) is an optimization problem to which (\ref%
{AmbiguityAverseMarketMaker2}) converges as the value of $\Omega $ increases
to infinity.

\begin{equation}
\begin{array}{c}
\max\limits_{\mathbf{x}}\min\limits_{\mathbf{p\in }\Psi
}\sum\limits_{i=1}^{N}p_{i}u\left[ \mathbf{b}^{T}\mathbf{x-}%
\sum_{j=1}^{J}A_{i,j}x_{j}\right] \\ 
\text{such that}\smallskip \\ 
\begin{tabular}{ll}
(A') & $\mathbf{0\leq x\leq Q}\smallskip $ \\ 
(B') & $\mathbf{p\geq 0},\sum_{i=1}^{N}p_{i}=1$ \\ 
(C') & $\Psi =\left\{ \mathbf{p\in 
\mathbb{R}
}^{N\times 1}|\mathbf{p\geq 0},\sum_{i=1}^{N}p_{i}=1,\sum_{i=1}^{N}p_{i}\ln
\left( \frac{p_{i}}{q_{i}}\right) \leq \Omega \right\} $%
\end{tabular}%
\end{array}
\label{AmbiguityAverseMarketMaker2}
\end{equation}

(\ref{AmbiguityAverseMarketMaker2}) is the optimization problem that the
market maker should be solving if the market maker's decision-making process
obeys the theory of Gilboa and Schmeidler (1989) or Ghirardato et al.
(2004). $\mathbf{b}^{T}\mathbf{x}$ is the total premium that the market
maker collects from the traders. $\sum_{j=1}^{J}A_{i,j}x_{j}$ is what the
market maker has to pay to traders if the $i$th outcome is realized. $%
\sum\limits_{i=1}^{N}p_{i}u\left[ \mathbf{b}^{T}\mathbf{x-}%
\sum_{j=1}^{J}A_{i,j}x_{j}\right] $ is thus the expected utility for the
market maker. The vector $\mathbf{q\in 
\mathbb{R}
}^{N}$ is the pivot prior probability distribution. The market maker
considers any probability distribution $\mathbf{p}$ reasonable as long as
the Kullback-Leibler distance between $\mathbf{p}$ and $\mathbf{q}$ is not
greater than $\Omega $.

Therefore, (\ref{AmbiguityAverseMarketMaker}) is an optimization problem
that the market maker solves if he/she is a DM with extreme Knightian
ambiguity aversion. In Theorem 2, we show that the market-clearing order
fill of the CPCAM is an optimal market-clearing strategy of a market maker
with extreme ambiguity aversion.

Our result may be relevant to other pari-mutuel markets because the CPCAM is
closely related to other pari-mutuel markets. First, the CPCAM is an
improved version of the PDCA. Peters et al. (2005) developed the CPCAM to
make the optimization problem convex. However, the CPCAM and the PDCA still
yield the same equilibrium price.

Second, Agrawal et al. (2011) show that many important pari-mutuel markets
in the literature (e.g., the Market Scoring Rule mechanism, cost-function
based market makers, utility-based market makers, and the Sequential Convex
Pari-mutuel Mechanism) can be understood under a common theoretical
framework. The Sequential Convex Pari-mutuel Mechanism (SCPM) (Peters et
al., 2007) is one of the pari-mutuel markets that Agrawal et al. (2011)
analyze. In addition, the CPCAM and the SCPM are very closely related to one
another. The only major difference is that the CPCAM is a call auction and
the SCPM\ is a continuous market. Therefore, the CPCAM and other important
pari-mutuel markets are closely related to one another. Given this close
relationship between different market designs, our analysis of the CPCAM may
also apply to other pari-mutuel markets. However, we leave that extension to
future work.

\section{The Knightian Pari-mutuel Mechanism (KPM)}

In this section, we design a new market called the Knightian Pari-mutuel
Mechanism (KPM).

\subsection{The Market Setting}

The limit order logic and the basic trading environment are similar to those
of the PDCA (Lange and Economide, 2005). However, the algorithm through
which the market maker clears the market is original. In particular, the
probabilistic treatment of the market maker's optimization problem is
original.

Like the CPCAM, the KPM allows the common market maker to handle multiple
types of contingent claims written on the same random event. There are $N$
possible states of the uncertain event, each of which is indexed by $i\in
\{1,2,...,N\}$.

The KPM allows traders to submit both market orders and limit orders.
Traders can submit market orders just as if they were submitting limit
orders simply by making the limit price extremely high or low. Therefore,
throughout the rest of the paper, we assume that people trade only limit
orders. When submitting each limit order, the trader indicates the limit
price, the limit quantity, and if the order is a buy or a sell.

For the sake of simplicity, we describe the setting in which the market is
run as a call auction. However, the setting can be easily adjusted to
accommodate continuous trading in the same manner as the CPCAM (Peters et
al., 2005) is changed to the SCPM (Peters et al., 2007).

Suppose there is a total of $J$ limit orders outstanding in the limit order
book. Let the matrix $\mathbf{A\in 
\mathbb{R}
}^{N\times J}$ represent the payoff structure of those orders. The column
matrix $\mathbf{A}_{\cdot j}\mathbf{\in 
\mathbb{R}
}^{N\times 1}$ is the payoff structure of the contingent claim that the $j$%
th order attempts to transact. For example, suppose that the second order
attempts to buy three shares of the contingent claim that pays \$1 per share
if and only if state 1 is realized. In such a case, the column matrix $%
\mathbf{A}_{\cdot 2}$ is $\left[ 
\begin{array}{ccccc}
1 & 0 & 0 & ... & 0%
\end{array}%
\right] ^{T}$.

When determining the market equilibrium price of each order, the market
maker first determines the equilibrium price for each state. Let $\mathbf{%
\xi =}\left[ 
\begin{array}{cccc}
\xi _{1} & \xi _{2} & ... & \xi _{N}%
\end{array}%
\right] ^{T}$ denote the equilibrium state prices. Then, the market maker
determines the market-clearing price of each contingent claim by taking the
dot product between the payoff vector and $\mathbf{\xi }$. For example,
consider the contingent claim with payoff structure $\left[ 
\begin{array}{cccc}
1 & 0 & 1 & 0%
\end{array}%
\right] ^{T}$. Suppose that the equilibrium state price vector $\mathbf{\xi }
$ is $\left[ 
\begin{array}{cccc}
\xi _{1} & \xi _{2} & \xi _{3} & \xi _{4}%
\end{array}%
\right] ^{T}$. Then, the market-clearing price of this contingent claim is $%
\xi _{1}+\xi _{3}$. Therefore, to determine the equilibrium price of each
order in the limit order book, the market maker only has to determine the
value of $\mathbf{\xi }$.

The binary variable $B_{j}$ is $1$ if the $j$th limit order is a buy order
and $-1$ if the $j$th limit order is a sell order. Let $b_{j}$ and $Q_{j}$
denote the limit price and the limit quantity associated with the $j$th
order, respectively. Let $x_{j}$ denote the actual number of the shares of
the claim that the submitter of the $j$th order is allowed to trade. We call 
$x_{j}$ the "order fill" for the $j$th order.

Once the equilibrium price of each order is determined, the market maker
decides $x_{j}$, $\forall j$ according to the limit order logic. Consider a
buy order. If the market-clearing price of an order is strictly higher than
the limit price, $x_{j}$ is exactly equal to $0$. If the market-clearing
price is strictly lower than the limit price, $x_{j}$ is set to $Q$. In
these two cases, the limit order logic automatically determines the order
fill. In contrast, if the market-clearing price of an order is exactly equal
to the limit price, the value of $x_{j}$ can be any number in the closed
interval $\left[ 0,Q_{j}\right] $. The logic works similarly for a sell
order. Define the vector $\mathbf{x\in 
\mathbb{R}
}^{J\times 1}$ such that the $j$th element of $\mathbf{x}$ is $x_{j}$.
Similarly, define $\mathbf{Q\in 
\mathbb{R}
}^{J\times 1}$ such that the $j$th element of $\mathbf{Q}$ is $Q_{j}$.
Define $\mathbf{b\in 
\mathbb{R}
}^{J\times 1}$ such that the $j$th element of $\mathbf{b}$ is $b_{j}$\textbf{%
.}

\begin{definition}[Limit Order Logic]
$\sum\limits_{i=1}^{N}A_{ij}\xi _{i}$ is the market-clearing price of the $j$%
th order. 
\begin{equation*}
\begin{array}{ccc}
x_{j}=0 & if & \sum\limits_{i=1}^{N}A_{ij}\xi _{i}>B_{j}b_{j} \\ 
x_{j}\in \left[ 0,Q_{j}\right] & if & \sum\limits_{i=1}^{N}A_{ij}\xi
_{i}=B_{j}b_{j} \\ 
x_{j}=Q_{j} & if & \sum\limits_{i=1}^{N}A_{ij}\xi _{i}<B_{j}b_{j}%
\end{array}%
\end{equation*}
\end{definition}

The market maker has two decision variables for his/her optimal clearing of
the market: the equilibrium state prices $\mathbf{\xi }$ and the order fill
vector $\mathbf{x}$.

Like other market makers in the financial markets, the market maker of the
KPM\ also has an inventory of contingent claims. If the $i$th state is
realized in the future, the inventory subjects the market maker to the
monetary payoff of $w_{i}$. Let $\alpha $ denote the market maker's risk
aversion coefficient. Suppose that the constant absolute risk aversion
(CARA) utility function $u(x)=-e^{-\alpha x}$ characterizes the market
maker's risk appetite.

The market maker has Knightian ambiguity toward the random future event on
which the claims are written. Let the set $\Psi $ define the set of
probability distributions that the market maker considers. $\mathbf{q\in 
\mathbb{R}
}^{N\times 1}$ is the market maker's pivot probability distribution. Assume
that every element in $\mathbf{q}$ is strictly positive. Any probability
distribution for which the Kullback-Leibler from $\mathbf{q}$ is no greater
than $\Omega $ is acceptable for the market maker. $\Omega $ quantifies the
market maker's level of ambiguity aversion. A large value of $\Omega $
implies that the market maker has strong ambiguity aversion. The $i$th
elements of $\mathbf{p}$ and $\mathbf{q}$ describe the market maker's
probabilistic belief about the $i$th outcome.

\begin{equation}
\Psi =\left\{ \mathbf{p\in 
\mathbb{R}
}^{N\times 1}|\mathbf{p\geq 0},\sum_{i=1}^{N}p_{i}=1,\sum_{i=1}^{N}p_{i}\ln
\left( \frac{p_{i}}{q_{i}}\right) \leq \Omega \right\}
\label{setofCandidates}
\end{equation}

\subsection{The Market-Clearing Optimization Problem}

We assume that the market maker adheres to the standard decision-making
theory under Knightian ambiguity aversion. The market maker's optimization
problem can be framed as (\ref{mainOptimization}). Unlike the CPCAM, the KPM
asks the market participants to pay the market-clearing prices of the claims
instead of the bid prices they submitted.

This optimization problem does not make any arbitrary assumptions. The
problem is a corollary of the standard theory of decision making under
ambiguity aversion. However, the constraints (E1) - (E3) and the objective
function causes the problem to be non-convex. Finding a global optimal
solution to a non-convex optimization problem is extremely difficult.

\begin{equation}
\begin{array}{c}
\max\limits_{\mathbf{\xi ,x}}\min\limits_{\mathbf{p\in }\Psi
}-\sum_{i=1}^{N}p_{i}\exp \left[ -\alpha w_{i}-\alpha
\sum\nolimits_{j=1}^{J}x_{j}\left( \left( \mathbf{A}^{T}\mathbf{\xi }\right)
_{j}-A_{ij}\right) \right] \\ 
\text{such that}\smallskip \\ 
\begin{tabular}{ll}
(A) & $\Psi =\left\{ \mathbf{p\in 
\mathbb{R}
}^{N\times 1}|\mathbf{p\geq 0},\sum_{i=1}^{N}p_{i}=1,\sum_{i=1}^{N}p_{i}\ln
\left( \frac{p_{i}}{q_{i}}\right) \leq \Omega \right\} $ \\ 
(B) & $\mathbf{\xi \geq 0}\smallskip $ \\ 
(C) & $\sum_{i=1}^{N}\xi _{i}=1\smallskip $ \\ 
(E1) & $\forall j\in \{1,2,..,J\},%
\begin{array}{ccc}
x_{j}=0 & \text{ if } & \left( \mathbf{A}^{T}\mathbf{\xi }\right)
_{j}>B_{j}b_{j}%
\end{array}%
$ \\ 
(E2) & $\forall j\in \{1,2,..,J\},%
\begin{array}{ccc}
x_{j}\in \left[ 0,Q_{j}\right] & \text{ if } & \left( \mathbf{A}^{T}\mathbf{%
\xi }\right) _{j}=B_{j}b_{j}%
\end{array}%
$ \\ 
(E3) & $\forall j\in \{1,2,..,J\},%
\begin{array}{ccc}
x_{j}=Q_{j} & \text{ if } & \left( \mathbf{A}^{T}\mathbf{\xi }\right)
_{j}<B_{j}b_{j}%
\end{array}%
$%
\end{tabular}%
\end{array}
\label{mainOptimization}
\end{equation}

\begin{corollary}
Suppose that the market maker holds zero inventory: $w_{i}=0$ for $\forall i$%
. As the value of $\Omega $ increases to infinity, the KPM becomes
completely pari-mutuel. The market maker incurs no loss regardless of the
outcome.

\begin{proof}
See the Appendix.
\end{proof}
\end{corollary}

The KPM may not be completely pari-mutuel in the sense that the market maker
can lose money with positive probability. However, Corollary 1 shows that
the KPM subsumes a completely pari-mutuel market. By adjusting the value of $%
\Omega $, the market designer can fine-tune the extent to which the market
is close to being completely pari-mutuel. The larger the value of $\Omega $,
the more completely pari-mutuel the market becomes.

For example, consider increasing the value of $\Omega $. Problem (\ref%
{mainOptimization}) then models the auctioneer with a large level of
ambiguity aversion. The ambiguity-averse DM is very sensitive to the
worst-case scenario. Thus, the auctioneer clears the market such that he/she
performs moderately even in the worst-case scenario. In other words, the
auctioneer does not want to lose too much money even in the worst-case
scenario.\footnote{%
The cost of this strategy is that the market maker may not be able to make a
great deal of money on the upside.} In the extreme case in which $\Omega $
diverges to infinity, the auctioneer becomes so conservative that he/she
does not want to lose any money under any circumstances. The market should
become completely pari-mutuel.

\subsection{The Market-Clearing Algorithm}

Before further discussion, we introduce new notations: $z_{i}=-e^{-\alpha
w_{i}}$ and $\theta _{i}=q_{i}e^{\Omega }$ for each $i\in \{1,2,...,N\}$. In
addition, let $\mathbf{\tciFourier }$ be the set of pairs $\left( \mathbf{%
\xi ,x}\right) $ that satisfy the limit order logic constraints (E1), (E2),
and (E3).

\begin{lemma}
$\left( \mathbf{\xi ,x}\right) =\left( \mathbf{\xi }^{\ast }\mathbf{,x}%
^{\ast }\right) $ is an optimal solution to (\ref{mainOptimization}) if and
only if it is part of an optimal solution to (\ref{mainOptimizationVer2}). 
\begin{equation}
\begin{array}{c}
\min\limits_{\mathbf{\xi ,x,}\mu ,\mathbf{d,\zeta }}\mu \ln \left(
\sum_{i=1}^{N}\theta _{i}e^{-\frac{d_{i}}{\mu }}\right) \\ 
\text{such that}\smallskip \\ 
\begin{tabular}{ll}
(A) & $-d_{i}=-z_{i}e^{\zeta _{i}}\text{ for }\forall i$ \\ 
(B) & $\zeta _{i}\geq \alpha \sum\nolimits_{j=1}^{J}\left[ x_{j}\mathbf{A}%
_{ij}-Q_{j}\left( \mathbf{A}^{T}\mathbf{\xi }\right) _{j}-B_{j}b_{j}\left(
x_{j}-Q_{j}\right) \right] $ for $\forall i$ \\ 
(C) & $\zeta _{i}\geq $ $\alpha \sum\nolimits_{j=1}^{J}\left[ x_{j}\mathbf{A}%
_{ij}-B_{j}b_{j}x_{j}\right] $ for $\forall i$ \\ 
(F) & $\mu \geq 0$ \\ 
(G) & $\mathbf{\xi \geq 0}\smallskip $ \\ 
(H) & $\sum_{i=1}^{N}\xi _{i}=1\smallskip $ \\ 
(I) & $(\mathbf{x,\xi )\in \tciFourier }$%
\end{tabular}%
\end{array}
\label{mainOptimizationVer2}
\end{equation}

\begin{proof}
See the Appendix.
\end{proof}
\end{lemma}

It is difficult to directly apply well-known optimization algorithms (e.g.,
the interior-point method) to solve (\ref{mainOptimizationVer2}) because the
problem is non-convex. The problem is non-convex because $\mathbf{%
\tciFourier }$ is not a convex set.

Suppose $\mathbf{C}$ is a convex set of pairs $(\mathbf{x,\xi )}$. We define
another optimization problem (\ref{smallBlackProblem}). 
\begin{equation}
\begin{array}{c}
\min\limits_{\mathbf{\xi ,x,}\mu ,\mathbf{d,\zeta }}\mu \ln \left(
\sum_{i=1}^{N}\theta _{i}e^{-\frac{d_{i}}{\mu }}\right) \\ 
\text{such that}\smallskip \\ 
\begin{tabular}{ll}
(A) & $-d_{i}=-z_{i}e^{\zeta _{i}}\text{ for }\forall i$ \\ 
(B) & $\zeta _{i}\geq \alpha \sum\nolimits_{j=1}^{J}\left[ x_{j}\mathbf{A}%
_{ij}-Q_{j}\left( \mathbf{A}^{T}\mathbf{\xi }\right) _{j}-B_{j}b_{j}\left(
x_{j}-Q_{j}\right) \right] $ for $\forall i$ \\ 
(C) & $\zeta _{i}\geq $ $\alpha \sum\nolimits_{j=1}^{J}\left[ x_{j}\mathbf{A}%
_{ij}-B_{j}b_{j}x_{j}\right] $ for $\forall i$ \\ 
(F) & $\mu \geq 0$ \\ 
(G) & $(\mathbf{x,\xi )\in C}$%
\end{tabular}%
\end{array}
\label{smallBlackProblem}
\end{equation}

\begin{lemma}
The optimization problem (\ref{smallBlackProblem}) is a convex optimization
problem.

\begin{proof}
See the Appendix.
\end{proof}
\end{lemma}

Our general strategy is as follows. First, we express the set of pairs $(%
\mathbf{x,\xi )}$ that satisfies the constraints (G), (H) and (I) in (\ref%
{mainOptimizationVer2}) as a union of multiple convex sets $\mathbf{C}_{1},%
\mathbf{C}_{2},$...$,\mathbf{C}_{M}$. Second, we solve a convex optimization
problem (\ref{smallBlackProblem}) with $\mathbf{C}$ replaced with each $%
\mathbf{C}_{m}$, $m\in \{1,...,M\}$. Let $L_{m}$ denote the optimal value of
the objective function from solving the convex optimization problem (\ref%
{smallBlackProblem}) with $\mathbf{C=C}_{m}$. Let $(\mathbf{x}_{m}\mathbf{%
,\xi }_{m}\mathbf{)}$ denote the optimal solutions to those problems. Third,
we find $m^{\ast }=\arg \max\limits_{m}L_{m}$. $(\mathbf{x}_{m^{\ast }}%
\mathbf{,\xi }_{m^{\ast }}\mathbf{)}$ becomes the global optimal solution to
the main optimization problem (\ref{mainOptimizationVer2}). By Lemma 1, $(%
\mathbf{x}_{m^{\ast }}\mathbf{,\xi }_{m^{\ast }}\mathbf{)}$ is the global
optimal solution to (\ref{mainOptimization}).

\subsubsection{Partitioning of the Feasible Set}

Suppose that a total of $K$ types of contingent claims are traded in the
market. Let the vector $P_{k}\in 
\mathbb{R}
^{N}$ denote the payoff structure of the $k$th security ($1\leq k\leq K$).
For example, if the $i$th outcome is realized, the person holding the claim
receives $P_{k,i}$ per share from the market maker.

Because there are $J$ outstanding orders in the limit order book, there are $%
J$ limit prices. Let $n_{k\text{ }}$denote the number of distinct limit
prices associated with the $k$th security. If there are multiple orders with
the same limit price and the same security, only one is counted toward $%
n_{k} $. Sort those bid prices in ascending order. Let $B_{k}^{l}$ denote
the $l$th smallest limit price associated with the $k$th security.

\begin{example}
For the sake of simplicity, consider a market in which only Arrow-Debreu
securities are traded. Suppose that $N=5$. Suppose that there are $K=5$
different Arrow-Debreu securities, one for each state of the world. The $k$%
th Arrow-Debreu security pays \$1 per share to its holder if and only if the 
$k$th state is realized.\smallskip \newline
Five row vectors in (\ref{PVector}) show the payoff structures of
Arrow-Debreu securities. For example, the nonzero entry in the first element
of $P_{1}$ implies that the first security pays \$1 per share if the first
outcome is realized.\smallskip\ 
\begin{equation}
\begin{tabular}{l}
$P_{1}=\left[ 
\begin{array}{ccccc}
1 & 0 & 0 & 0 & 0%
\end{array}%
\right] \medskip $ \\ 
$P_{2}=\left[ 
\begin{array}{ccccc}
0 & 1 & 0 & 0 & 0%
\end{array}%
\right] \medskip $ \\ 
$P_{3}=\left[ 
\begin{array}{ccccc}
0 & 0 & 1 & 0 & 0%
\end{array}%
\right] \medskip $ \\ 
$P_{4}=\left[ 
\begin{array}{ccccc}
0 & 0 & 0 & 1 & 0%
\end{array}%
\right] \medskip $ \\ 
$P_{5}=\left[ 
\begin{array}{ccccc}
0 & 0 & 0 & 0 & 1%
\end{array}%
\right] $%
\end{tabular}
\label{PVector}
\end{equation}%
\smallskip \newline
Table 1 illustrates seven orders ($J=7$) outstanding in the limit order
book. For example, the person who submitted the first order wants to buy the
first Arrow-Debreu security. The entry 0.18 in the fourth column implies
that he/she is willing to pay at most 0.18 dollars per share. The payoff
matrix in the last five columns shows how the person will be paid by the
market maker. For example, the market maker will pay \$1 to the person who
submitted the first order if and only if the first outcome is realized. The
person who submitted the fourth order will receive \$1 if and only if the
fourth outcome is realized.\smallskip

\ 
\begin{table}[tbp] \centering%
\begin{tabular}{|c|c|c|c|c|c|c|c|c|}
\hline
order & limit & security & limit & \multicolumn{5}{|c|}{payoff matrix} \\ 
\cline{5-9}
& quantity &  & price & \multicolumn{5}{|c|}{outcome \#} \\ \cline{5-9}
\# & $\mathbf{Q}$ & \# & $\mathbf{b}$ & 1 & 2 & 3 & 4 & 5 \\ \hline\hline
1 & 0.001 & 1 & 0.18 & 1 & 0 & 0 & 0 & 0 \\ \hline
2 & 0.001 & 2 & 0.18 & 0 & 1 & 0 & 0 & 0 \\ \hline
3 & 0.001 & 3 & 0.18 & 0 & 0 & 1 & 0 & 0 \\ \hline
4 & 0.001 & 4 & 0.18 & 0 & 0 & 0 & 1 & 0 \\ \hline
5 & 0.002 & 1 & 0.20 & 1 & 0 & 0 & 0 & 0 \\ \hline
6 & 0.001 & 1 & 0.25 & 1 & 0 & 0 & 0 & 0 \\ \hline
7 & 0.001 & 1 & 0.20 & 1 & 0 & 0 & 0 & 0 \\ \hline
\end{tabular}%
\caption{An example of a limit order book}\label{TableKey}%
\end{table}%
\newline
Based on this limit order book, we can determine the number of distinct
limit prices associated with each Arrow-Debreu security. For example, for
the first Arrow-Debreu security, there are three distinct limit prices:
0.18, 0.20 and 0.25. Thus, $n_{1}$ should be 3. Likewise, $n_{2}=1$, $%
n_{3}=1 $, $n_{4}=1$, and $n_{5}=0$.\medskip \newline
Next, we sort the limit prices in ascending order. For example, for the
first Arrow-Debreu security, we have $B_{1}^{1}=0.18$, $B_{1}^{2}=0.20$, and 
$B_{1}^{3}=0.25$. In addition, $B_{2}^{1}=0.18$, $B_{3}^{1}=0.18$, and $%
B_{4}^{1}=0.18$. Because there is no limit order associated with the fifth
Arrow-Debreu security, $B_{5}^{1}$ is undefined. $\blacksquare $
\end{example}

Suppose that there are $n_{k}$ distinct limit prices. Let $E$ denote the $N$%
-dimensional space defined as (\ref{setEdefinition}). We define $n_{k}+1$
convex subsets of $E$ such that if $\mathbf{\xi }$ is restricted to one of
those subsets, (\ref{mainOptimization}) becomes a convex optimization
problem. The intuition is as follows. The limit order logic constraints (E1)
- (E3) are non-convex because we do not know which of the three conditions - 
$\left( \mathbf{A}^{T}\mathbf{\xi }\right) _{j}>b_{j}$ or $\left( \mathbf{A}%
^{T}\mathbf{\xi }\right) _{j}=b_{j}$ or $\left( \mathbf{A}^{T}\mathbf{\xi }%
\right) _{j}<b_{j}$ - hold at an optimal solution. We define subsets to
ensure that such ambiguitiy is resolved within each set. As a result, the
limit order logic constraints can be replaced by $x_{j}=0$ or $x_{j}\in %
\left[ 0,Q_{j}\right] $ or $x_{j}=Q_{j}$.%
\begin{equation}
E=\left\{ \mathbf{\xi \in 
\mathbb{R}
}^{N}|\sum\limits_{i=1}^{N}\xi _{i}=1,\mathbf{\xi \geq 0}\right\}
\label{setEdefinition}
\end{equation}

Let us illustrate how we obtain subsets of $E$. Note that the
market-clearing price of each Arrow-Debreu security is bounded below by 0
and above by 1. $n_{k\text{ }}$distinct bid prices associated with the $k$th
Arrow-Debreu security define $2n_{k}+1$ subsets of $[0,1]$: $[0,B_{k}^{1}]$, 
$B_{k}^{1}$, $[B_{k}^{1},B_{k}^{2}]$, $B_{k}^{2}$,...,$B_{k}^{n_{k}}$,$%
[B_{k}^{n_{k}},1]$.\footnote{%
I assume that the bid prices are strictly larger than 0 and strictly smaller
than 1.} These $2n_{k}+1$ points or closed intervals can be used to define $%
2n_{k}+1$ subsets of $E$: $E_{k}^{1}$, $E_{k}^{2}$,..., $E_{k}^{2n_{k}+1}$,
as shown in (\ref{partitionEExample}).

\begin{equation}
\begin{tabular}{c}
$E_{k}^{1}=\left\{ \mathbf{\xi \in 
\mathbb{R}
}^{N}|\sum\limits_{i=1}^{N}\xi _{i}=1,\mathbf{\xi \geq 0,}P_{k}\mathbf{\xi
\in }[0,B_{k}^{1}]\right\} $ \\ 
$E_{k}^{2}=\left\{ \mathbf{\xi \in 
\mathbb{R}
}^{N}|\sum\limits_{i=1}^{N}\xi _{i}=1,\mathbf{\xi \geq 0,}P_{k}\mathbf{\xi =}%
B_{k}^{1}\right\} $ \\ 
... \\ 
$E_{k}^{2n_{k}+1}=\left\{ \mathbf{\xi \in 
\mathbb{R}
}^{N}|\sum\limits_{i=1}^{N}\xi _{i}=1,\mathbf{\xi \geq 0,}P_{k}\mathbf{\xi
\in }[B_{k}^{n_{k}},1]\right\} $%
\end{tabular}
\label{partitionExample}
\end{equation}

\begin{example}
We continue with the earlier example. Let us begin with the first
Arrow-Debreu security. Using the three distinct limit prices, we can define $%
2\times 3+1=7$ subsets of $E=\left\{ \mathbf{\xi \in 
\mathbb{R}
}^{5}|\sum\limits_{i=1}^{5}\xi _{i}=1,\mathbf{\xi \geq 0}\right\} $: $%
E_{1}^{1},E_{1}^{2},...,E_{1}^{7}$. Note that $P_{1}\mathbf{\xi =}\xi _{1}$.%
\newline
\begin{equation}
\begin{tabular}{l}
$E_{1}^{1}=\left\{ \mathbf{\xi \in 
\mathbb{R}
}^{5}|\sum\limits_{i=1}^{5}\xi _{i}=1,\mathbf{\xi \geq 0,}0\leq \xi _{1}\leq
0.18\right\} $ \\ 
$E_{1}^{2}=\left\{ \mathbf{\xi \in 
\mathbb{R}
}^{5}|\sum\limits_{i=1}^{5}\xi _{i}=1,\mathbf{\xi \geq 0,}\xi
_{1}=0.18\right\} $ \\ 
$E_{1}^{3}=\left\{ \mathbf{\xi \in 
\mathbb{R}
}^{5}|\sum\limits_{i=1}^{5}\xi _{i}=1,\mathbf{\xi \geq 0,}0.18\leq \xi
_{1}\leq 0.2\right\} $ \\ 
$E_{1}^{4}=\left\{ \mathbf{\xi \in 
\mathbb{R}
}^{5}|\sum\limits_{i=1}^{5}\xi _{i}=1,\mathbf{\xi \geq 0,}\xi
_{1}=0.2\right\} $ \\ 
$E_{1}^{5}=\left\{ \mathbf{\xi \in 
\mathbb{R}
}^{5}|\sum\limits_{i=1}^{5}\xi _{i}=1,\mathbf{\xi \geq 0,}0.2\leq \xi
_{1}\leq 0.25\right\} $ \\ 
$E_{1}^{6}=\left\{ \mathbf{\xi \in 
\mathbb{R}
}^{5}|\sum\limits_{i=1}^{5}\xi _{i}=1,\mathbf{\xi \geq 0,}\xi
_{1}=0.25\right\} $ \\ 
$E_{1}^{7}=\left\{ \mathbf{\xi \in 
\mathbb{R}
}^{5}|\sum\limits_{i=1}^{5}\xi _{i}=1,\mathbf{\xi \geq 0,}0.25\leq \xi
_{1}\leq 1\right\} $%
\end{tabular}
\label{partitionEExample}
\end{equation}%
There is only one distinct limit price for the second Arrow-Debreu security.
Therefore, we can define three subsets of the set $E$ as (\ref%
{partitionEExample2}). The third and the fourth Arrow-Debreu securities also
have only one limit price. Therefore, the partitioning of $E$ should work in
exactly the same way. 
\begin{equation}
\begin{tabular}{l}
$E_{2}^{1}=\left\{ \mathbf{\xi \in 
\mathbb{R}
}^{5}|\sum\limits_{i=1}^{5}\xi _{i}=1,\mathbf{\xi \geq 0,}0\leq \xi _{2}\leq
0.18\right\} $ \\ 
$E_{2}^{2}=\left\{ \mathbf{\xi \in 
\mathbb{R}
}^{5}|\sum\limits_{i=1}^{5}\xi _{i}=1,\mathbf{\xi \geq 0,}\xi
_{2}=0.18\right\} $ \\ 
$E_{2}^{3}=\left\{ \mathbf{\xi \in 
\mathbb{R}
}^{5}|\sum\limits_{i=1}^{5}\xi _{i}=1,\mathbf{\xi \geq 0,}0.18\leq \xi
_{2}\leq 1\right\} $%
\end{tabular}
\label{partitionEExample2}
\end{equation}%
There is no outstanding order or limit price associated with the fifth
Arrow-Debreu security. Thus, we can define only one subset of the set $E$: $%
E_{5}^{1}$. 
\begin{equation}
E_{5}^{1}=\left\{ \mathbf{\xi \in 
\mathbb{R}
}^{5}|\sum\limits_{i=1}^{5}\xi _{i}=1,\mathbf{\xi \geq 0}\right\}
\label{partitionExample3}
\end{equation}%
$\blacksquare $
\end{example}

We introduce new notation as (\ref{intersectionsE}). The idea is as follows.
Having $n_{1\text{ }}$distinct limit prices associated with the first
security yields $2n_{1}+1$ distinct subsets of $E$. Choose one subset out of
these $2n_{1}+1$ subsets. Similarly, choose one of $2n_{2}+1$ subsets of $E$
that we generate from the limit prices associated with the second security.
Repeat this process for the remaining Arrow-Debreu securities. Once we have
one subset for each type of Arrow-Debreu security, we can obtain the
intersection of those $K$ subsets, which is shown in (\ref{intersectionsE}).
There are a total of $\Pi _{k=1}^{K}(2n_{k}+1)$ ways to choose a combination
of subsets. 
\begin{equation}
\begin{tabular}{c}
$E(\ell _{1},\ell _{2},...,\ell _{K})=E_{1}^{\ell _{1}}\cap E_{2}^{\ell
_{2}}\cap \text{...}\cap E_{K}^{\ell _{K}}$ \\ 
where $1\leq \ell _{1}\leq 2n_{1}+1,...,1\leq \ell _{K}\leq 2n_{K}+1$%
\end{tabular}
\label{intersectionsE}
\end{equation}

Now consider the optimization problem (\ref{mainOptimizationVer2}). Imagine
replacing the constraint $\sum\limits_{i=1}^{N}\xi _{i}=1,\mathbf{\xi \geq 0}
$ with a more restrictive one (\ref{intersectionsE}). The part that causes
problem (\ref{mainOptimizationVer2}) to be non-convex is (\ref{troubleMaker}%
). However, once the feasible set of the state price vector $\mathbf{\xi }$
is restricted to a smaller set $E(\ell _{1},\ell _{2},...,\ell _{K})$, (\ref%
{troubleMaker}) can be replaced with $x_{j}=0$ or $x_{j}\in \left[ 0,Q_{j}%
\right] $ or $x_{j}=Q_{j}$ for $\forall j$.%
\begin{equation}
\begin{array}{c}
\begin{array}{ccc}
x_{j}=0 & \text{ if } & \left( \mathbf{A}^{T}\mathbf{\xi }\right)
_{j}>B_{j}b_{j}%
\end{array}
\\ 
\begin{array}{ccc}
x_{j}\in \left[ 0,Q_{j}\right] & \text{ if } & \left( \mathbf{A}^{T}\mathbf{%
\xi }\right) _{j}=B_{j}b_{j}%
\end{array}
\\ 
\begin{array}{ccc}
x_{j}=Q_{j} & \text{ if } & \left( \mathbf{A}^{T}\mathbf{\xi }\right)
_{j}<B_{j}b_{j}%
\end{array}%
\end{array}%
\text{for }\forall j\in \{1,...,J\}  \label{troubleMaker}
\end{equation}

\begin{example}
Again, we continue with the previous example. Because $n_{1}=3$, $n_{2}=1$, $%
n_{3}=1$, $n_{4}=1$, and $n_{5}=0$ there are in total $(3\times 2+1)\times
(1\times 2+1)\times (1\times 2+1)\times (1\times 2+1)\times (0\times
2+1)=189 $ different sets of the form $E(\ell _{1},\ell _{2},\ell _{3},\ell
_{4},\ell _{5})$. \newline
To illustrate, consider a particular case in which $\ell _{1}=1,\ell
_{2}=2,\ell _{3}=2,\ell _{4}=2,$ and $\ell _{5}=1$. 
\begin{equation}
\begin{tabular}{l}
$E_{1}^{\ell _{1}}=\left\{ \mathbf{\xi \in 
\mathbb{R}
}^{5}|\sum\limits_{i=1}^{5}\xi _{i}=1,\mathbf{\xi \geq 0,}0\leq \xi _{0}\leq
0.18\right\} $ \\ 
$E_{2}^{\ell _{2}}=\left\{ \mathbf{\xi \in 
\mathbb{R}
}^{5}|\sum\limits_{i=1}^{5}\xi _{i}=1,\mathbf{\xi \geq 0,}\xi
_{1}=0.18\right\} $ \\ 
$E_{3}^{\ell _{3}}=\left\{ \mathbf{\xi \in 
\mathbb{R}
}^{5}|\sum\limits_{i=1}^{5}\xi _{i}=1,\mathbf{\xi \geq 0,}\xi
_{2}=0.18\right\} $ \\ 
$E_{4}^{\ell _{4}}=\left\{ \mathbf{\xi \in 
\mathbb{R}
}^{5}|\sum\limits_{i=1}^{5}\xi _{i}=1,\mathbf{\xi \geq 0,}\xi
_{3}=0.18\right\} $ \\ 
$E_{5}^{\ell _{5}}=\left\{ \mathbf{\xi \in 
\mathbb{R}
}^{5}|\sum\limits_{i=1}^{5}\xi _{i}=1,\mathbf{\xi \geq 0}\right\} $%
\end{tabular}
\label{exampleEl}
\end{equation}%
\begin{eqnarray}
E(\ell _{1} &=&1,\ell _{2}=2,\ell _{3}=2,\ell _{4}=2,\ell _{5}=1)
\label{exampleIntersectionE} \\
&=&\left\{ \mathbf{\xi \in 
\mathbb{R}
}^{5}|\sum\limits_{i=1}^{5}\xi _{i}=1,\mathbf{\xi \geq 0,}0\leq \xi _{1}\leq
0.18,\xi _{2}=\xi _{3}=\xi _{4}=0.18\right\}  \notag
\end{eqnarray}%
Suppose that we replace the usual constraint $\sum\limits_{i=1}^{N}\xi
_{i}=1,\mathbf{\xi \geq 0}$ with a more restrictive one (\ref%
{exampleIntersectionE}) in the main optimization problem. Then, the
optimization should take the form of (\ref{moreRestrictiveOptimization}). 
\begin{equation}
\begin{array}{c}
\min\limits_{\mathbf{\xi ,x,}\mu \mathbf{,\zeta ,\omega }}\mu \ln \left(
\sum_{i=1}^{5}\theta _{i}e^{-\frac{d_{i}}{\mu }}\right) \\ 
\text{such that}\smallskip \\ 
\begin{tabular}{ll}
(A) & $\omega _{i}\geq -z_{i}e^{\zeta _{i}}\text{ for }\forall i$ \\ 
(B) & $\zeta _{i}\geq \alpha \sum\nolimits_{j=1}^{7}\left[ x_{j}\mathbf{A}%
_{ij}-Q_{j}\left( \mathbf{A}^{T}\mathbf{\xi }\right) _{j}-B_{j}b_{j}\left(
x_{j}-Q_{j}\right) \right] $, $\forall i$ \\ 
(C) & $\zeta _{i}\geq \alpha \sum\nolimits_{j=1}^{7}\left[ x_{j}\mathbf{A}%
_{ij}-B_{j}b_{j}x_{j}\right] $, $\forall i$ \\ 
(D) & $\mu \geq 0$ \\ 
(E) & $\mathbf{\xi \in }$ $E(1,2,2,2,1)$ \\ 
(F) & $%
\begin{array}{c}
\begin{array}{ccc}
x_{j}=0 & \text{ if } & \left( \mathbf{A}^{T}\mathbf{\xi }\right)
_{j}>B_{j}b_{j}%
\end{array}
\\ 
\begin{array}{ccc}
x_{j}\in \left[ 0,Q_{j}\right] & \text{ if } & \left( \mathbf{A}^{T}\mathbf{%
\xi }\right) _{j}=B_{j}b_{j}%
\end{array}
\\ 
\begin{array}{ccc}
x_{j}=Q_{j} & \text{ if } & \left( \mathbf{A}^{T}\mathbf{\xi }\right)
_{j}<B_{j}b_{j}%
\end{array}%
\end{array}%
$for $\forall j$%
\end{tabular}%
\end{array}
\label{moreRestrictiveOptimization}
\end{equation}%
As long as constraint (E) holds, constraint (F) can be replaced with those
in the last column of Table 2. 
\begin{eqnarray*}
&&%
\begin{tabular}{|c|c|c|c||c|c|c|}
\hline
order & limit & security & bid & market & relevant & restriction \\ 
\# & quantity & \# & price & clearing & restriction & on the \\ 
& $\mathbf{b}$ &  &  & price & in $E(1,2,2,2,1)$ & order fill $x_{j}$ \\ 
\hline\hline
1 & 0.001 & 1 & 0.18 & $\xi _{1}$ & $\xi _{1}\leq 0.18$ & $x_{1}=0.001$ \\ 
\hline
2 & 0.001 & 2 & 0.18 & $\xi _{2}$ & $\xi _{2}=0.18$ & $0\leq x_{2}\leq 0.001$
\\ \hline
3 & 0.001 & 3 & 0.18 & $\xi _{3}$ & $\xi _{3}=0.18$ & $0\leq x_{3}\leq 0.001$
\\ \hline
4 & 0.001 & 4 & 0.18 & $\xi _{4}$ & $\xi _{4}=0.18$ & $0\leq x_{4}\leq 0.001$
\\ \hline
5 & 0.002 & 1 & 0.20 & $\xi _{1}$ & $\xi _{1}\leq 0.18$ & $x_{5}=0.001$ \\ 
\hline
6 & 0.001 & 1 & 0.25 & $\xi _{1}$ & $\xi _{1}\leq 0.18$ & $x_{6}=0.001$ \\ 
\hline
7 & 0.001 & 1 & 0.20 & $\xi _{1}$ & $\xi _{1}\leq 0.18$ & $x_{7}=0.001$ \\ 
\hline
\end{tabular}
\\
&&\text{Table 2 An Example of How the Limit Order Logic Constraint Can be
Simplified }
\end{eqnarray*}%
Solving (\ref{moreRestrictiveOptimization}) is equivalent to solving (\ref%
{smallerOptimizationEx}).%
\begin{equation}
\begin{array}{c}
\min\limits_{\mathbf{\xi ,x,}\mu \mathbf{,\zeta ,\omega }}\ell \left( \mu
\right) =\mu \ln \left( \sum_{i=1}^{5}\theta _{i}e^{-\frac{d_{i}}{\mu }%
}\right) \\ 
\text{such that}\smallskip \\ 
\begin{tabular}{ll}
(A) & $-d_{i}\geq -z_{i}e^{\zeta _{i}}\text{ for }\forall i$ \\ 
(B) & $\zeta _{i}\geq \alpha \sum\nolimits_{j=1}^{7}\left[ x_{j}\mathbf{A}%
_{ij}-Q_{j}\left( \mathbf{A}^{T}\mathbf{\xi }\right) _{j}-B_{j}b_{j}\left(
x_{j}-Q_{j}\right) \right] $, $\forall i$ \\ 
(C) & $\zeta _{i}\geq \alpha \sum\nolimits_{j=1}^{7}\left[ x_{j}\mathbf{A}%
_{ij}-B_{j}b_{j}x_{j}\right] $, $\forall i$ \\ 
(D) & $\mu \geq 0$ \\ 
(E) & $\mathbf{\xi \in }$ $E(1,2,2,2,1)$ \\ 
(F) & $x_{1}=x_{5}=x_{6}=x_{7}=0.001$, $0\leq x_{2},x_{3},x_{4}\leq 0.001$%
\end{tabular}%
\end{array}
\label{smallerOptimizationEx}
\end{equation}%
$\blacksquare $
\end{example}

For notational simplicity, we define new sets:%
\begin{eqnarray}
&&X(\ell _{1},\ell _{2},...,\ell _{m})  \label{defineSetX} \\
&=&\left\{ \mathbf{x\in 
\mathbb{R}
}^{J}|%
\begin{tabular}{lll}
$x_{j}=0$ & if & $\max\limits_{\mathbf{\xi \in }E(\ell _{1},\ell
_{2},...,\ell _{K})}\left( \mathbf{A}^{T}\mathbf{\xi }\right)
_{j}>B_{j}b_{j} $ and \\ 
$x_{j}\in \left[ 0,Q_{j}\right] $ & if & $\min\limits_{\mathbf{\xi \in }%
E(\ell _{1},\ell _{2},...,\ell _{K})}\left( \mathbf{A}^{T}\mathbf{\xi }%
\right) _{j}=\max\limits_{\mathbf{\xi \in }E(\ell _{1},\ell _{2},...,\ell
_{K})}\left( \mathbf{A}^{T}\mathbf{\xi }\right) _{j}=B_{j}b_{j}$ \\ 
$x_{j}=Q_{j}$ & if & $\min\limits_{\mathbf{\xi \in }E(\ell _{1},\ell
_{2},...,\ell _{K})}\left( \mathbf{A}^{T}\mathbf{\xi }\right)
_{j}<B_{j}b_{j} $%
\end{tabular}%
\text{, }\forall j\right\}  \notag
\end{eqnarray}

\begin{example}
We continue with the previous example. $X(1,2,2,2,1)$ is defined as (\ref%
{exampleOfXDefinition}).%
\begin{equation}
X(1,2,2,2,1)=\left\{ \mathbf{x\in 
\mathbb{R}
}^{7}|x_{1}=x_{5}=x_{6}=x_{7}=0.001,0\leq x_{2},x_{3},x_{4}\leq 0.001\right\}
\label{exampleOfXDefinition}
\end{equation}
\end{example}

\subsubsection{The Pseudo-Code}

If we\ apply an interative method (e.g., the interior point method) to solve
(\ref{smallBlackProblem}), $\mu $ may converge toward zero along the path.
However, the objective function is ill-defined when $\mu $ is zero.
Therefore, we define a new objective function as (\ref{newObjectiveL}). 
\begin{equation}
L\left( \mu ,\mathbf{\omega }\right) =%
\begin{array}{ccc}
\mu \ln \left( \sum_{i=1}^{N}\theta _{i}e^{\frac{\omega _{i}}{\mu }}\right)
& if & \mu >0 \\ 
\max_{1\leq i\leq N}\omega _{i} & if & \mu =0%
\end{array}
\label{newObjectiveL}
\end{equation}

Then, (\ref{mainOptimizationVer2})\ can be reformulated as (\ref%
{mainOptimizationVer3}). 
\begin{equation}
\begin{array}{c}
\min\limits_{\mathbf{\xi ,x,}\mu \mathbf{,\zeta ,\omega }}L\left( \mu ,%
\mathbf{\omega }\right) \\ 
\text{such that}\smallskip \\ 
\begin{tabular}{ll}
(A) & $\omega _{i}\geq -z_{i}e^{\zeta _{i}}\text{ for }\forall i$ \\ 
(B) & $\zeta _{i}\geq \alpha \sum\nolimits_{j=1}^{J}\left[ x_{j}\mathbf{A}%
_{ij}-Q_{j}\left( \mathbf{A}^{T}\mathbf{\xi }\right) _{j}-B_{j}b_{j}\left(
x_{j}-Q_{j}\right) \right] $, $\forall i$ \\ 
(C) & $\zeta _{i}\geq \alpha \sum\nolimits_{j=1}^{J}\left[ x_{j}\mathbf{A}%
_{ij}-B_{j}b_{j}x_{j}\right] $, $\forall i$ \\ 
(D) & $\mu \geq 0$ \\ 
(E) & $\mathbf{\xi \geq 0}\smallskip $ \\ 
(F) & $\sum\limits_{i=1}^{N}\xi _{i}=1$ \\ 
(G) & $(\mathbf{x,\xi )\in \tciFourier }$%
\end{tabular}%
\end{array}
\label{mainOptimizationVer3}
\end{equation}

A global optimal solution to (\ref{mainOptimizationVer3}) can be obtained by
executing the following pseudo-code. 
\begin{equation}
\begin{array}{c}
\min\limits_{\mathbf{\xi ,x,}\mu \mathbf{,\zeta ,\omega }}L\left( \mu ,%
\mathbf{\omega }\right) \\ 
\text{such that}\smallskip \\ 
\begin{tabular}{ll}
(A) & $\omega _{i}\geq -z_{i}e^{\zeta _{i}}\text{ for }\forall i$ \\ 
(B) & $\zeta _{i}\geq \alpha \sum\nolimits_{j=1}^{J}\left[ x_{j}\mathbf{A}%
_{ij}-Q_{j}\left( \mathbf{A}^{T}\mathbf{\xi }\right) _{j}-B_{j}b_{j}\left(
x_{j}-Q_{j}\right) \right] $, $\forall i$ \\ 
(C) & $\zeta _{i}\geq \alpha \sum\nolimits_{j=1}^{J}\left[ x_{j}\mathbf{A}%
_{ij}-B_{j}b_{j}x_{j}\right] $, $\forall i$ \\ 
(D) & $\mu \geq 0$ \\ 
(E) & $\mathbf{x\in }$ $E(\ell _{1},\ell _{2},...,\ell _{K})$ \\ 
(F) & $\mathbf{\xi \in }$ $X(\ell _{1},\ell _{2},...,\ell _{K})$%
\end{tabular}%
\end{array}
\label{pseudoCodeSubProblem}
\end{equation}

\begin{equation}
\begin{tabular}{l}
for $\ell _{1}=1:1:n_{1}$ \\ 
\ \ \ \ \ \ for $\ell _{2}=1:1:n_{2}$ \\ 
\ \ \ \ \ \ \ \ \ ... \\ 
\ \ \ \ \ \ \ \ \ \ \ \ \ for $\ell _{K}=1:1:n_{K}$ \\ 
\ \ \ \ \ \ \ \ \ \ \ \ \ \ \ \ \ \ \ \ \ if $E(\ell _{1},\ell _{2},...,\ell
_{K})\neq \varnothing $ \\ 
\ \ \ \ \ \ \ \ \ \ \ \ \ \ \ \ \ \ \ \ \ \ \ \ \ \ Solve (\ref%
{pseudoCodeSubProblem}) using the interior point method. \\ 
\ \ \ \ \ \ \ \ \ \ \ \ \ \ \ \ \ \ \ \ \ \ \ \ \ \ The optimal value of the
objective function $\rightarrow $ $L^{\ast }(\ell _{1},\ell _{2},...,\ell
_{K})$ \\ 
\ \ \ \ \ \ \ \ \ \ \ \ \ \ \ \ \ \ \ \ \ \ \ \ \ \ The optimizing value of $%
\mathbf{x}$ $\rightarrow $ $\mathbf{x}^{\ast }(\ell _{1},\ell _{2},...,\ell
_{K})$ \\ 
\ \ \ \ \ \ \ \ \ \ \ \ \ \ \ \ \ \ \ \ \ \ \ \ \ \ The optimizing value of $%
\mathbf{\xi }$ $\rightarrow $ $\mathbf{\xi }^{\ast }(\ell _{1},\ell
_{2},...,\ell _{K})$ \\ 
\ \ \ \ \ \ \ \ \ \ \ \ \ \ \ \ \ \ \ \ \ end \\ 
\ \ \ \ \ \ \ \ \ \ \ \ \ end \\ 
\ \ \ \ \ \ \ \ \ ... \\ 
\ \ \ \ \ \ for \\ 
end \\ 
$\arg \max_{\ell _{1},\ell _{2},...,\ell _{K},E(\ell _{1},\ell _{2},...,\ell
_{K})\neq \varnothing }L^{\ast }(\ell _{1},\ell _{2},...,\ell _{K})$ $%
\rightarrow $ $\ell _{1}^{\ast },\ell _{2}^{\ast },...,\ell _{K}^{\ast }$ \\ 
$\mathbf{x}^{\ast }(\ell _{1},\ell _{2},...,\ell _{K})$, $\mathbf{\xi }%
^{\ast }(\ell _{1},\ell _{2},...,\ell _{K})$ $\rightarrow $ global optimal
solution%
\end{tabular}
\label{PseudoCoe}
\end{equation}

\subsubsection{The Computational Efficiency}

In modern complexity analysis, the efficiency of an algorithm is assessed
based on whether the number of iterations required is bounded above by a
polynomial of the problem dimension (Luenberger and Ye, 2008). In our
setting, the number of securities traded in the market typically does not
grow in the order of thousands. Frequently, a growing number of outstanding
orders in the limit order book demands significant computing power.
Therefore, to prove that our algorithm is of practical value, we need to
show that the algorithm is polynomial in the number of outstanding orders $J$%
. Theorem 3 does precisely this.

\begin{theorem}
The number of iterations required to execute the pseudo-code (\ref{PseudoCoe}%
) is bounded above by\ a polynomial function of the number of outstanding
orders $J$.
\end{theorem}

\begin{proof}
See the Appendix.
\end{proof}

\section{Simulation}

For simplicity, we simulate the market when only Arrow-Debreu securities are
traded. The $i$th security pays \$1 per share to the holder if and only if
state $i$ is realized at maturity, where $i\in \{1,2,3,4,5\}$.

Through this simulation exercise, we verify that our market-clearing
algorithm gives the result that is consistent with economic intuition.

\subsection{Simulation A: Market Maker's Ambiguity Aversion}

In this subsection, we present simulation results that show how the market
maker's level of ambiguity aversion affects how the market is cleared. We
run simulations for five different parameters of ambiguity aversion: $\Omega
=0,0.2,0.4,1,2$. Table 3 shows the sample limit order book used for
Simulation A. Table 4 summarizes the parameters used for each of the five
iterations. The table reports the number of shares that traders as a whole
hold. Thus, any positive number in the top-left part of the table implies
that the market maker may have to incur additional loss at the time the
securities mature.

\begin{eqnarray*}
&&%
\begin{tabular}{|c|c|c|c|c|c|c|c|c|c|}
\hline
order & limit & security & bid price & \multicolumn{5}{|c|}{payoff matrix} & 
buy \\ 
\# & quantity & \# & per & state & state & state & state & state & or \\ 
& $\mathbf{b}$ &  & share & 1 & 2 & 3 & 4 & 5 & sell \\ \hline\hline
1 & 0.002 & 1 & 0.18 & 1 & 0 & 0 & 0 & 0 & buy \\ \hline
2 & 0.001 & 2 & 0.18 & 0 & 1 & 0 & 0 & 0 & buy \\ \hline
3 & 0.001 & 3 & 0.18 & 0 & 0 & 1 & 0 & 0 & buy \\ \hline
4 & 0.001 & 4 & 0.18 & 0 & 0 & 0 & 1 & 0 & buy \\ \hline
5 & 0.001 & 5 & 0.18 & 0 & 0 & 0 & 0 & 1 & buy \\ \hline
\end{tabular}
\\
&&\text{Table 3 Sample Limit Order Book Used for Simulation A}
\end{eqnarray*}%
\begin{equation*}
\FRAME{itbpFU}{5.3186in}{3.0597in}{0in}{\Qcb{Table 4: The Set of Parameters
Used for Each Iteration}}{}{simulationtable1.bmp}{\special{language
"Scientific Word";type "GRAPHIC";maintain-aspect-ratio TRUE;display
"USEDEF";valid_file "F";width 5.3186in;height 3.0597in;depth
0in;original-width 15.4862in;original-height 8.8747in;cropleft "0";croptop
"1";cropright "1";cropbottom "0";filename
'simulationTable1.bmp';file-properties "XNPEU";}}
\end{equation*}

In Table 3, we purposefully make the bid prices slightly lower than 0.2. For
example, because of the limit order logic, the market-clearing price of the
first state has to be equal to or smaller than 0.18 for the market maker to
accept the first order. If he/she wants to accept all five outstanding
orders, he/she has to make every state price lower than or equal to 0.18.
However, because the state prices must sum to 1, it is impossible to do so.
Therefore, the market maker has to strategically accept some orders while
declining others. We investigate how the market maker's ambiguity aversion
affects this strategic decision making through this simulation.

When making a strategic choice over the five outstanding orders, there are
two counteracting forces. The first factor derives from the skewness in the
pivot prior probability distribution. This factor causes the market maker to
want to accept orders \#1, \#2, or \#3. According to the market maker's
subjective probabilistic belief, he/she is very unlikely to be forced to pay
money at the time the securities mature. However, the factor causes the
market maker to not want to accept order \#4 or \#5.

This factor becomes weaker with an increasingly large value of $\Omega $.
Suppose that the value of $\Omega $ becomes increasingly large. The set $%
\Psi $ of probability distributions that the market maker considers in
his/her decision making becomes larger. As a result, the upper bound and the
lower bound on the probability of a particular outcome becomes higher and
lower, respectively. The widening gap between the upper and lower bounds
causes the market maker's probabilistic belief to be increasingly
uninformative. For example, suppose that the value of $\Omega $ is extremely
large. The probability of a particular outcome can be as high as 1 and as
low as 0. In such a case, it is as if the market maker had no information
about the event. In conclusion, a larger value of $\Omega $ causes the
market maker to further disregard the pivot prior distribution in making the
decision, thereby weakening the first factor.

The second factor derives from the market maker's aversion to extreme
downside risk. The price of any Arrow-Debreu security is between 0 and 1.
Thus, the worst-case payoff of any Arrow-Debreu security is typically
negative for the market maker.\footnote{%
It is zero if and only if the price is 1.} Fearing this worst-case scenario,
the ambiguity-averse market maker will not want to purchase the security.
This factor causes the market maker to not want to fill any outstanding
order.

The second factor becomes stronger with an increasing value of $\Omega $. $%
\Omega $ is a parameter that captures the extent to which the market maker
is ambiguity averse. The larger the value of $\Omega $, the more ambiguity
averse the market maker. Ambiguity aversion causes the DM to become obsessed
with the worst-case scenario. Therefore, a large value of $\Omega $ causes
the second factor to become stronger.

Figure 1 shows the simulation result. The result can be easily interpreted
using the two counteracting forces we just explained. First, consider the
situation in which $\Omega $\ is small. The first factor dominates the
second factor. As a result, the market maker accepts orders \#1, \#2 and \#3
while declining orders \#4 and \#5. Second, consider the case in which $%
\Omega $\ is large. Here, the second factor dominates the first factor. The
market maker does not accept any order when $\Omega $ is larger than 0.4.

\FRAME{ftbpFU}{2.6887in}{2.7726in}{0pt}{\Qcb{The graph shows the order fills
for different values of $\Omega $, which parametrizes the market maker's
level of ambiguity aversion. For example, when $\Omega $ is 0.4, a 0.001
share of the second Arrow-Debreu security is filled. The second Arrow-Debreu
security pays \$1 to its holder if and only if the second state is realized
at maturity.}}{}{simulation_omegaeffect_notitle.bmp}{\special{language
"Scientific Word";type "GRAPHIC";maintain-aspect-ratio TRUE;display
"USEDEF";valid_file "F";width 2.6887in;height 2.7726in;depth
0pt;original-width 7.9027in;original-height 8.1526in;cropleft "0";croptop
"1";cropright "1";cropbottom "0";filename
'simulation_omegaEffect_NoTitle.bmp';file-properties "XNPEU";}}

\subsection{Simulation B: Market Maker's Pivot Probability Distribution}

In this subsection, we\ show how the market maker's pivot prior
probabilistic belief affects the way our algorithm clears the market. Table
5 below shows the limit order book used for this subsection. Table 6 below
shows the set of simulation parameters for both iterations.%
\begin{eqnarray*}
&&%
\begin{tabular}{|c|c|c|c|c|c|c|c|c|c|}
\hline
order & limit & security & bid price & \multicolumn{5}{|c|}{payoff matrix} & 
buy \\ 
\# & quantity & \# & per & state & state & state & state & state & or \\ 
& $\mathbf{b}$ &  & share & 1 & 2 & 3 & 4 & 5 & sell \\ \hline\hline
1 & 0.001 & 1 & 0.18 & 1 & 0 & 0 & 0 & 0 & buy \\ \hline
2 & 0.001 & 2 & 0.18 & 0 & 1 & 0 & 0 & 0 & buy \\ \hline
3 & 0.001 & 3 & 0.18 & 0 & 0 & 1 & 0 & 0 & buy \\ \hline
4 & 0.001 & 4 & 0.18 & 0 & 0 & 0 & 1 & 0 & buy \\ \hline
5 & 0.001 & 5 & 0.18 & 0 & 0 & 0 & 0 & 1 & buy \\ \hline
\end{tabular}
\\
&&\text{Table 5 Sample Limit Order Book Used for Simulation B}
\end{eqnarray*}

\begin{equation*}
\FRAME{itbpFU}{5.8678in}{2.3194in}{0in}{\Qcb{Table 6: The Set of Parameters
Used for Simulation B}}{}{simulationtable2.bmp}{\special{language
"Scientific Word";type "GRAPHIC";maintain-aspect-ratio TRUE;display
"USEDEF";valid_file "F";width 5.8678in;height 2.3194in;depth
0in;original-width 15.4862in;original-height 6.0831in;cropleft "0";croptop
"1";cropright "1";cropbottom "0";filename
'simulationTable2.bmp';file-properties "XNPEU";}}
\end{equation*}

Figure 2 shows the two prior distributions used in this simulation exercise.
The market maker with the uniform prior has no information on what will
happen in the future. With no valuable piece of evidence available to make
an inference, the market maker simply assumes that each state is equally
probable. In contrast, the market maker with the exponential prior is more
assertive in deciding which state is more probable than the others. For
example, he/she thinks that state 5 is at least sixty times more probable
than state 1.

\FRAME{ftbpFU}{4.1917in}{2.7726in}{0pt}{\Qcb{The two prior distributions
used for the simulation. For example, the market maker with the exponential
prior believes that state 5 will be realized with 63.6\% probability.}}{}{%
simulation_twopriordistributions_notitle.bmp}{\special{language "Scientific
Word";type "GRAPHIC";maintain-aspect-ratio TRUE;display "USEDEF";valid_file
"F";width 4.1917in;height 2.7726in;depth 0pt;original-width
10.6804in;original-height 7.0413in;cropleft "0";croptop "1";cropright
"1";cropbottom "0";filename
'simulation_twoPriordistributions_NoTitle.bmp';file-properties "XNPEU";}}

Figure 3 shows the outcome of the market-clearing algorithm for two prior
distributions. We interpret the result with the countervailing forces
introduced in the previous section.

\FRAME{ftbpFU}{2.6091in}{2.7in}{0pt}{\Qcb{The graph shows the market
clearance result for different prior beliefs held by the market maker. The
vertical axis shows the number of shares of each Arrow-Debreu security
filled. For example, in the case of the exponential prior, 0.001 shares of
the first Arrow-Debreu security are filled (the first Arrow-Debreu security
pays \$1 to its holder if and only if state 1 is realized at maturity). }}{}{%
simulation_priorbeliefeffect_notitle.bmp}{\special{language "Scientific
Word";type "GRAPHIC";maintain-aspect-ratio TRUE;display "USEDEF";valid_file
"F";width 2.6091in;height 2.7in;depth 0pt;original-width
7.5282in;original-height 7.792in;cropleft "0";croptop "1";cropright
"1";cropbottom "0";filename
'simulation_priorBeliefEffect_NoTitle.bmp';file-properties "XNPEU";}}

For the market maker with the completely uninformative prior, the second
factor strongly dominates the first factor. The first factor is powerless
because, with the uniform prior, the market maker does not face a larger
risk of incurring a loss in one state than in the other states. Dominated by
the second force, the market maker does not accept any order.

On the contrary, the first factor is much stronger for the market maker with
the exponential prior. The first, the second, and the third states receive
very small probability weights. Therefore, the market maker views selling
the first security as an opportunity to make a riskless profit of 0.18
dollars per share. By a similar line of reasoning, the market maker has a
strong disincentive against accepting the fifth order. The result is in
accordance with this intuition. Figure 3 shows that the market maker with
the exponential prior accepts only the first, second, and third orders.

\section{The Strength of the KPM: Empirical Discussion}

Based on a solid understanding of a pari-mutuel auctioneer from the
perspective of market microstructure theory, we discuss why a market making
firm may want to organize a derivative market based on the KPM.

\subsection{Why Automate?}

The KPM is an automated market maker. The main strength of an automated
market maker relative to its human counterpart derives from its ability to
update quotes for dozens of related securities almost instantaneously. This
ability reduces adverse selection cost, thereby allowing market makers to
provide more competitive quotes to customers.

In today's increasingly electronic and automated trading environment, the
market maker's ability to quickly update his/her quotes is increasingly
important. The various speeds with which market participants react to the
arrival of new information represent a source of informational asymmetry
(Foucault et al., 2003; Litzenberger, 2012). In particular, liquidity
suppliers who are slow to react to new information can leave their stale
quotes vulnerable to being adversely picked off by high-frequency traders
(Hendershott and Riordan, 2013). The competition to respond to new
information faster than anyone else has become so intense that trading firms
want to place their computers the building where the exchange's matching
machine is: The time it takes for the light to travel from their computers
to the matching machine matters (Litzenberger, 2012). Given this
extraordinarily high-frequency trading environment, the automation of the
quote-updating process is important for the liquidity supplier to survive.

This adverse selection cost becomes particularly important for a market
maker involved in multiple related markets: Quotes need to be consistent
with one another to ensure that there is no arbitrage opportunity. With more
information to process, the comparative advantage of the automated market
maker over the human counterpart can only become more significant (Gerig and
Michayluk, 2013).

The KPM\ is an automated algorithm through which the liquidity supplier can
quickly price multiple contingent claims while taking into account a variety
of factors. The resulting prices reflect the market maker's risk aversion
and ambiguity aversion while ensuring that there is no arbitrage opportunity.

\subsection{Other Well-Known Strengths of a Pari-mutuel Auction}

First, the ISE is interested in the PDCA\ mainly because pari-mutuel markets
can effectively mitigate counterparty risk (Burne, 2013). The pari-mutuel
auctioneer can be thought of as the central clearing counterparty (CCP). In
particular, the pari-mutuel auctioneer is the common CCP operating in
multiple contingent claims markets. The fact that one auctioneer handles
multiple markets allows the pari-mutuel market to better mitigate
counterparty risk.\footnote{%
Duffie and Zhu (2011) show that counterparty risk can be better managed if
the same CCP is involved in more than one market.}

Second, the auction performs better than other trading platforms,
particularly in a low-liquidity environment. The auction aggregates
liquidity dispersed in multiple individual markets into the common pool.
Lastly, having the common market maker in multiple markets improves price
efficiency (Lange and Economide, 2005). Please see the Appendix for further
details.

\subsection{Potential Areas of Application}

The KPM is expected to be useful for options markets in which delta hedging
the market maker's inventory is not feasible.\footnote{%
Baron and Lange (2007) also argue that the PDCA\ is suitable for markets
where delta hedging is difficult.} The KPM solves an optimization that is
robust to worst-case scenarios. In particular, Corollary 1 shows that the
KPM can become almost completely pari-mutuel when the value of $\Omega $ is
very large. If the market is completely pari-mutuel, the market maker does
not lose any money regardless of what happens at maturity. Therefore, the
inability to delta hedge the inventory becomes less critical.

There are two specific options markets for which delta hedging may be
particularly infeasible. The first example is options for which the
underlying asset is not tradable (e.g., the market for economic derivatives
written on U.S. non-farm payrolls) (Baron and Lange, 2007). The second
example is options with extremely short time to maturity because the delta
fluctuates too much (Baron and Lange, 2007).

\section{Conclusion}

In this paper, we first show that the market-clearing strategy of the Convex
Pari-mutuel Call Auction Mechanism (CCPAM) is asymptotically equivalent to
that of the market maker with extreme ambiguity aversion for the future
contingent event. Because the CPCAM is closely related to other notable
pari-mutuel auctions in the literature, we regard this conclusion as a basis
for arguing that pari-mutuel auctions are closely related to ambiguity
aversion.

With this understanding, we design a new market for trading contingent
claims, the Knightian Pari-mutuel Mechanism (KPM). The main optimization
problem of the KPM is what the market maker should solve if he/she adheres
to the theory of decision making under ambiguity aversion. The algorithm
clears the market while controlling for the market maker's level of risk and
ambiguity aversion. We present a polynomial-time algorithm to solve the
optimization problem.

Our paper may contribute to facilitating the adoption of a pari-mutuel
mechanism in the trading community. As Robert Shiller once noted, a
pari-mutuel mechanism can be particularly useful in launching a wide variety
of innovative derivatives markets, thereby enabling investors to hedge a new
class of fundamental risks (Baron and Lange, 2007).

\section{Appendix}

\subsection{Illustration of the Theory of Decision Making Under Uncertainty}

Suppose that there is an urn that contains red, blue and green balls, of
which there are 90 in total. While there are 30 red balls in the urn, the
exact number of either blue balls or green balls is unknown to the DM.
Suppose that five lotteries are available. Lottery R pays \$1 if and only if
the DM draws a red ball from the urn. Lotteries B and G pay \$1 if and only
if he/she draws a blue ball and a green ball, respectively. Similarly,
lottery RB pays \$1 if and only if either a red ball or a blue ball is
drawn. Lottery BG pays \$1 if and only if either a blue ball or a green ball
is drawn. Empirical studies show that most people prefer lottery R to either
lottery B or G. Moreover, most people prefer lottery BG to RB. It is well
known that this empirical result contradicts Savage's theory of utility
maximization with subjective probability (Savage, 1954).

Let us reformulate the DM's problem in the language of Theorem 1. The set of
all possible states $S$ is $\left\{ red,blue,green\right\} $. The set of
consequences $X$ is $\{0,1\}$, expressed in dollars. The DM\ is interested
in five different acts: $f_{R}$, $f_{B}$, $f_{G}$, $f_{RB}$, and $f_{BG}$.\
The act $f_{R}:S\longrightarrow X$ is a mapping such that $f_{R}(red)=1$, $%
f_{R}(blue)=0$ and $f_{R}(green)=0$. We define the other four acts similarly.

Without knowing the exact number of either blue or green balls, the DM
cannot attach a single probability distribution to $S$. Suppose that the
DM's set of candidates is $\Psi =\{(1/3,x,2/3-x)\in 
\mathbb{R}
^{3}|0.1\leq x\leq 0.4\},$ where $1/3,x,$and $2/3-x$ are the chances of
drawing red, blue and green balls, respectively.

Let $u:X\longrightarrow 
\mathbb{R}
$ denote the DM's utility function. Equations (\ref{ellsbergExample1}) and (%
\ref{ellsbergExample2}) should hold for the DM to prefer lottery R to the
other two lotteries.\medskip 
\begin{subequations}
\begin{equation}
\min\limits_{0.1\leq x\leq 0.4}\left[ \frac{1}{3}u(1)+\frac{2}{3}u(0)\right]
\geq \min\limits_{0.1\leq x\leq 0.4}\left[ x\cdot u(1)+(1-x)\cdot u(0)\right]
\label{ellsbergExample1}
\end{equation}%
\begin{equation}
\min\limits_{0.1\leq x\leq 0.4}\left[ \frac{1}{3}u(1)+\frac{2}{3}u(0)\right]
\geq \min\limits_{0.1\leq x\leq 0.4}\left[ \left( \frac{2}{3}-x\right) \cdot
u(1)+(\frac{1}{3}+x)\cdot u(0)\right] \medskip  \label{ellsbergExample2}
\end{equation}%
In addition, equation (\ref{ellsbergExample3}) must hold for the DM to
prefer lottery BG to RB. 
\end{subequations}
\begin{equation}
\min\limits_{0.1\leq x\leq 0.4}\left[ \frac{2}{3}u(1)+\frac{1}{3}u(0)\right]
\geq \min\limits_{0.1\leq x\leq 0.4}\left[ \left( \frac{1}{3}+x\right) \cdot
u(1)+(\frac{2}{3}-x)\cdot u(0)\right]  \label{ellsbergExample3}
\end{equation}%
Equations (\ref{ellsbergExample1}), (\ref{ellsbergExample2}) and (\ref%
{ellsbergExample3}) hold as long as the utility function is non-decreasing.
Theorem 1 successfully reconciles the theory with empirical observations. $%
\blacksquare $

\subsection{The Proof of Theorem 2}

If we\ assume that $\delta _{i}=\delta $ for $\forall i$, (\ref{CPCAM}) is a
barrier problem to (\ref{CPCAM2}). 
\begin{equation}
\begin{array}{c}
\max\limits_{\mathbf{x,}M}\mathbf{b}^{T}\mathbf{x-}M \\ 
\text{such that}\smallskip \\ 
\begin{tabular}{ll}
(A) & $\sum_{j=1}^{J}A_{i,j}x_{j}\leq M$ for each $i\in \{1,2,...,N\}$ \\ 
(B) & $\mathbf{0\leq x\leq Q}\smallskip $%
\end{tabular}%
\end{array}
\label{CPCAM2}
\end{equation}

Assume that the feasible set for (\ref{CPCAM2}) is not empty. Sending the
value of the parameter $\delta $ to zero is equivalent to reducing the
duality gap along the primal-dual central path in the interior point method.
Thus, as $\delta $ approaches zero, $\mathbf{x}(\delta )$ should converge to
an optimal solution to (\ref{CPCAM2}) (Luenberger and Ye, 2008), which we
denote $\mathbf{x}^{\ast }$.

(\ref{CPCAM2}) is equivalent to (\ref{CPCAM3}). 
\begin{equation}
\begin{array}{c}
\max\limits_{\mathbf{x}}\left[ \mathbf{b}^{T}\mathbf{x-}\max\limits_{i}%
\sum_{j=1}^{J}A_{i,j}x_{j}\right] \\ 
\text{such that}\smallskip \\ 
\begin{tabular}{ll}
(A') & $\mathbf{0\leq x\leq Q}\smallskip $%
\end{tabular}%
\end{array}
\label{CPCAM3}
\end{equation}

(\ref{CPCAM3}) is equivalent to (\ref{CPCAM4'}).%
\begin{equation}
\begin{array}{c}
\max\limits_{\mathbf{x}}\min\limits_{i}u\left[ \mathbf{b}^{T}\mathbf{x-}%
\sum_{j=1}^{J}A_{i,j}x_{j}\right] \\ 
\text{such that}\smallskip \\ 
\begin{tabular}{ll}
(A') & $\mathbf{0\leq x\leq Q}\smallskip $%
\end{tabular}%
\end{array}
\label{CPCAM4'}
\end{equation}

(\ref{CPCAM4'}) is equivalent to (\ref{CPCAM5}).%
\begin{equation}
\begin{array}{c}
\max\limits_{\mathbf{x}}\min\limits_{\mathbf{p}}\sum\limits_{i=1}^{N}p_{i}u%
\left[ \mathbf{b}^{T}\mathbf{x-}\sum_{j=1}^{J}A_{i,j}x_{j}\right] \\ 
\text{such that}\smallskip \\ 
\begin{tabular}{ll}
(A') & $\mathbf{0\leq x\leq Q}\smallskip $ \\ 
(B') & $\mathbf{p\geq 0},\sum_{i=1}^{N}p_{i}=1$%
\end{tabular}%
\end{array}
\label{CPCAM5}
\end{equation}

\subsection{The Proof of Corollary 1}

With the assumption that $w_{i}=0$ for $\forall i$, (\ref{mainOptimization})
converges to (\ref{MainOptimizationSpecial1}) as the value of $\Omega $
increases to infinity.

\begin{equation}
\begin{array}{c}
\max\limits_{\mathbf{\xi ,x}}\min\limits_{\mathbf{p}}-\sum_{i=1}^{N}p_{i}%
\exp \left[ -\alpha \sum\nolimits_{j=1}^{J}x_{j}\left( \left( \mathbf{A}^{T}%
\mathbf{\xi }\right) _{j}-A_{ij}\right) \right] \\ 
\text{such that}\smallskip \\ 
\begin{tabular}{ll}
(A) & $\mathbf{p\geq 0},\sum_{i=1}^{N}p_{i}=1$ \\ 
(B) & $\mathbf{\xi \geq 0}\smallskip $ \\ 
(C) & $\sum_{i=1}^{N}\xi _{i}=1\smallskip $ \\ 
(E1) & $\forall j\in \{1,2,..,J\},%
\begin{array}{ccc}
x_{j}=0 & \text{ if } & \left( \mathbf{A}^{T}\mathbf{\xi }\right)
_{j}>B_{j}b_{j}%
\end{array}%
$ \\ 
(E2) & $\forall j\in \{1,2,..,J\},%
\begin{array}{ccc}
x_{j}\in \left[ 0,Q_{j}\right] & \text{ if } & \left( \mathbf{A}^{T}\mathbf{%
\xi }\right) _{j}=B_{j}b_{j}%
\end{array}%
$ \\ 
(E3) & $\forall j\in \{1,2,..,J\},%
\begin{array}{ccc}
x_{j}=Q_{j} & \text{ if } & \left( \mathbf{A}^{T}\mathbf{\xi }\right)
_{j}<B_{j}b_{j}%
\end{array}%
$%
\end{tabular}%
\end{array}
\label{MainOptimizationSpecial1}
\end{equation}

Let $\mathbf{\xi }^{\ast }$ and $\mathbf{x}^{\ast }$ denote the values of $%
\mathbf{\xi }$ and $\mathbf{x}$ that optimize (\ref{MainOptimizationSpecial1}%
), respectively. Define $i^{\ast }$ as (\ref{iAsterik}). $i^{\ast }$ may not
be uniquely defined. In that case, we simply choose any of multiple $i$s
that minimize $-\exp \left[ -\alpha \sum\nolimits_{j=1}^{J}x_{j}\left(
\left( \mathbf{A}^{T}\mathbf{\xi }\right) _{j}-A_{ij}\right) \right] $.%
\begin{equation}
i^{\ast }=\arg \min\limits_{i}-\exp \left[ -\alpha
\sum\nolimits_{j=1}^{J}x_{j}^{\ast }\left( \left( \mathbf{A}^{T}\mathbf{\xi }%
^{\ast }\right) _{j}-A_{ij}\right) \right]  \label{iAsterik}
\end{equation}

Consider the inner minimization problem. To minimize the objective function,
we need $p_{i^{\ast }}=1$ and $p_{i}=0$ for $\forall i\neq i^{\ast }$. If
we\ substitute $p_{i^{\ast }}=1$ and $p_{i}=0$ for $\forall i\neq i^{\ast }$
into the objective function of (\ref{MainOptimizationSpecial1}), we obtain (%
\ref{objectiveFunctionSpecial}). 
\begin{equation}
-\exp \left[ -\alpha \sum\nolimits_{j=1}^{J}x_{j}^{\ast }\left( \left( 
\mathbf{A}^{T}\mathbf{\xi }^{\ast }\right) _{j}-A_{i^{\ast }j}\right) \right]
=\min\limits_{i}-\exp \left[ -\alpha \sum\nolimits_{j=1}^{J}x_{j}^{\ast
}\left( \left( \mathbf{A}^{T}\mathbf{\xi }^{\ast }\right) _{j}-A_{ij}\right) %
\right]  \label{objectiveFunctionSpecial}
\end{equation}

If we substitute $\mathbf{x=0}$ into the objective function of (\ref%
{MainOptimizationSpecial1}), we obtain 1. Therefore, the optimal value of
the objective function of (\ref{MainOptimizationSpecial1}), which is (\ref%
{objectiveFunctionSpecial}), should be at least as large as 1. 
\begin{equation}
\min\limits_{i}-\exp \left[ -\alpha \sum\nolimits_{j=1}^{J}x_{j}^{\ast
}\left( \left( \mathbf{A}^{T}\mathbf{\xi }^{\ast }\right) _{j}-A_{ij}\right) %
\right] \geq 1  \label{inequality}
\end{equation}

(\ref{inequality}) is equivalent to (\ref{inequality2}).

\begin{equation}
\min\limits_{i}\sum\nolimits_{j=1}^{J}x_{j}^{\ast }\left( \left( \mathbf{A}%
^{T}\mathbf{\xi }^{\ast }\right) _{j}-A_{ij}\right) \geq 0
\label{inequality2}
\end{equation}

$\sum\nolimits_{j=1}^{J}x_{j}^{\ast }\left( \left( \mathbf{A}^{T}\mathbf{\xi 
}^{\ast }\right) _{j}-A_{ij}\right) $ is the monetary payoff for the market
maker if the $i$th outcome is realized. Therefore, $\min\limits_{i}\sum%
\nolimits_{j=1}^{J}x_{j}^{\ast }\left( \left( \mathbf{A}^{T}\mathbf{\xi }%
^{\ast }\right) _{j}-A_{ij}\right) $ is the worst possible monetary payoff
that the market maker can ever receive. Inequality (\ref{inequality2}) shows
that the market maker never loses money even in that worst-case scenario.
The market is completely pari-mutuel.

\subsection{The Proof of Lemma 1}

\subsubsection{The Dual Problem of the Inner Minimization Problem}

(\ref{firstConvex}) is the inner minimization problem isolated from (\ref%
{mainOptimization}). The optimal value of this inner optimization problem is
an implicit function of $\mathbf{\xi }$ and $\mathbf{x}$. Because the
objective function is linear in $\mathbf{p}$ and $\Psi $ is a convex set,
this problem is a convex optimization problem. 
\begin{equation}
\begin{array}{c}
\min\limits_{\mathbf{p\in }\Psi }\sum_{i=1}^{N}p_{i}z_{i}\exp \left[ -\alpha
\sum\nolimits_{j=1}^{J}x_{j}\left( \left( \mathbf{A}^{T}\mathbf{\xi }\right)
_{j}-A_{ij}\right) \right] \\ 
\text{such that}\smallskip \\ 
\begin{tabular}{ll}
(A) & $\Psi =\left\{ \mathbf{p\in 
\mathbb{R}
}^{N\times 1}|\mathbf{p\geq 0},\sum_{i=1}^{N}p_{i}=1,\sum_{i=1}^{N}p_{i}\ln
\left( \frac{p_{i}}{q_{i}}\right) \leq \Omega \right\} $%
\end{tabular}%
\end{array}
\label{firstConvex}
\end{equation}

$\mathbf{x}$ and $\mathbf{\xi }$ should be treated like constants when
solving (\ref{firstConvex}). To make notations simpler, we introduce new
constants.

\begin{equation}
d_{i}=z_{i}\exp \left[ -\alpha \sum\nolimits_{j=1}^{J}x_{j}\left( \left( 
\mathbf{A}^{T}\mathbf{\xi }\right) _{j}-A_{ij}\right) \right] \text{ for }%
\forall i\in \left\{ 1,..,N\right\}  \label{defineD}
\end{equation}%
\begin{equation}
\mathbf{d=}\left( 
\begin{array}{cccc}
d_{1} & d_{2} & ... & d_{N}%
\end{array}%
\right) ^{T}  \label{defineDVector}
\end{equation}

Then, minimization problem (\ref{firstConvex}) reduces to (\ref%
{firstConvexSimpler}).%
\begin{equation}
\begin{array}{c}
\min\limits_{\mathbf{p}}\mathbf{d}^{T}\mathbf{p} \\ 
\text{such that}\smallskip \\ 
\begin{tabular}{ll}
(A) & $\mathbf{p\geq 0}$ \\ 
(B) & $\sum_{i=1}^{N}p_{i}=1$ \\ 
(C) & $\sum_{i=1}^{N}p_{i}\ln \left( \frac{p_{i}}{q_{i}}\right) \leq \Omega $%
\end{tabular}%
\end{array}
\label{firstConvexSimpler}
\end{equation}

The domain of the minimization problem is $D$ as defined in (\ref%
{defineDomain}).%
\begin{equation}
D=\left\{ \mathbf{p\in 
\mathbb{R}
}^{N}|\mathbf{p>0}\right\}  \label{defineDomain}
\end{equation}

The Lagrangian associated with problem (\ref{firstConvexSimpler}) is (\ref%
{LagrangianFirstProblem}). $\lambda _{1}$, $\lambda _{2}$,..., $\lambda _{N}$%
, $\mu $, $\nu $ are Lagrange multipliers. 
\begin{equation}
L(\mathbf{p,\lambda ,}\mu ,\nu )=\mathbf{d}^{T}\mathbf{p}+\sum_{i=1}^{N}%
\lambda _{i}\left( -p_{i}\right) +\mu \left\{ \sum_{i=1}^{N}p_{i}\ln \left( 
\frac{p_{i}}{q_{i}}\right) -\bar{\Omega}\right\} +\nu \left(
\sum_{i=1}^{N}p_{i}-1\right)  \label{LagrangianFirstProblem}
\end{equation}

The Lagrange dual function associated with problem (\ref{firstConvexSimpler}%
) is (\ref{lagrangeDualFirstProb}).%
\begin{equation}
g(\mathbf{\lambda ,}\mu ,\nu )=\inf\limits_{\mathbf{p}>0}L(\mathbf{p,\lambda
,}\mu ,\nu )  \label{lagrangeDualFirstProb}
\end{equation}

$L(\mathbf{p,\lambda ,}\mu ,\nu )$ is a convex function of each $p_{i}$. The
first order condition is%
\begin{equation}
\frac{\partial L(\mathbf{p,\lambda ,}\mu ,\nu )}{\partial p_{i}}=\left(
d_{i}-\lambda _{i}+\nu \right) +\mu \left\{ 1+\ln \left( \frac{p_{i}}{q_{i}}%
\right) \right\} =0  \label{FOCFirstProblem}
\end{equation}%
\begin{equation}
1+\ln \left( \frac{p_{i}}{q_{i}}\right) =-\frac{d_{i}-\lambda _{i}+\nu }{\mu 
}  \label{FOCFirstProblem2}
\end{equation}%
\begin{equation}
p_{i}=q_{i}\exp \left( -1+\frac{\lambda _{i}-d_{i}-\nu }{\mu }\right) >0
\label{FOCFirstProblem3}
\end{equation}

Because the function is convex, (\ref{FOCFirstProblem3}) is the global
minimizer. We substitute (\ref{FOCFirstProblem3}) into (\ref%
{lagrangeDualFirstProb}). 
\begin{eqnarray}
&&g(\mathbf{\lambda ,}\mu ,\nu )  \notag \\
&=&\sum_{i=1}^{N}\left( d_{i}-\lambda _{i}\right) q_{i}e^{-1+\frac{\lambda
_{i}-d_{i}-\nu }{\mu }}+\sum_{i=1}^{N}\left( -\mu +\lambda _{i}-d_{i}-\nu
\right) q_{i}e^{-1+\frac{\lambda _{i}-d_{i}-\nu }{\mu }}-\mu \Omega
+v\sum_{i=1}^{N}q_{i}e^{-1+\frac{\lambda _{i}-d_{i}-\nu }{\mu }}-v  \notag \\
&=&\sum_{i=1}^{N}\left( -\mu \right) q_{i}e^{-1+\frac{\lambda _{i}-d_{i}-\nu 
}{\mu }}-\mu \Omega -\nu  \notag \\
&=&-\mu \sum_{i=1}^{N}q_{i}e^{-1+\frac{\lambda _{i}-d_{i}-\nu }{\mu }}-\mu
\Omega -\nu
\end{eqnarray}

The Lagrange dual problem associated with the inner minimization problem is (%
\ref{lagrangeDualProblemFirstMin}).%
\begin{equation}
\begin{array}{c}
\max\limits_{\mathbf{\lambda ,}\mu ,\nu }-\mu \sum_{i=1}^{N}q_{i}e^{-1+\frac{%
\lambda _{i}-d_{i}-\nu }{\mu }}-\mu \Omega -\nu \\ 
s.t. \\ 
\lambda _{1},...,\lambda _{N}\geq 0 \\ 
\mu \geq 0%
\end{array}
\label{lagrangeDualProblemFirstMin}
\end{equation}

Each $q_{i}$ is assumed to be positive. $\mu $ is implicity assumed to be
nonzero because it appears as a denominator in (\ref%
{lagrangeDualProblemFirstMin}). Therefore, the objective function of (\ref%
{lagrangeDualProblemFirstMin}) decreases with increasing $\lambda _{i}$. The
optimal value of each $\lambda _{i}$ should thus be zero. (\ref%
{lagrangeDualProblemFirstMin}) reduces to (\ref{lagraugeDualProblemFirstMin2}%
).%
\begin{equation}
\begin{array}{c}
\max\limits_{\mu ,\nu }-\mu \sum_{i=1}^{N}q_{i}e^{-1-\frac{d_{i}+\nu }{\mu }%
}-\mu \Omega -\nu \\ 
s.t. \\ 
\mu \geq 0%
\end{array}
\label{lagraugeDualProblemFirstMin2}
\end{equation}

\subsubsection{Applying Strong Duality to the Inner Minimization Problem}

We use the trick presented in Palomar (2009) to address the max-min problem.
We\ replace the inner minimization problem with the dual maximization
problem. This substitution is valid if and only if strong duality holds.
Then, the overall structure of the problem is max-max instead of max-min.
The double max structure can collapse to a more conventional problem with
only one maximization operator.

We use the criteria in Boyd and Vandenberghe (2004) to determine whether
strong duality holds. If the primal problem is convex and Slater's condition
holds, strong duality holds. Slater's condition holds if there exists a
strictly feasible $\mathbf{p\in relintD}$.

Slater's condition holds in the context of our problem as long as $\Omega $
is a strictly positive number (i.e., $\mathbf{p}$ such that $p_{i}=q_{i}$
for $\forall i$ is a strictly feasible solution). Therefore, strong duality
holds for our inner minimization problem as long as $\Omega $ is strictly
positive.

Using strong duality, our max-min problem can be transformed into (\ref%
{maxMaxVersion}).%
\begin{equation}
\begin{array}{c}
\max\limits_{\mathbf{\xi ,x,d}}\max\limits_{\mu ,\nu }-\mu
\sum_{i=1}^{N}q_{i}e^{-1-\frac{d_{i}+\nu }{\mu }}-\mu \Omega -\nu \\ 
\text{such that}\smallskip \\ 
\begin{tabular}{ll}
(A) & $d_{i}=z_{i}\exp \left[ -\alpha \sum\nolimits_{j=1}^{J}x_{j}\left(
\left( \mathbf{A}^{T}\mathbf{\xi }\right) _{j}-\mathbf{A}_{ij}\right) \right]
\text{ for }\forall i\in \left\{ 1,..,N\right\} $ \\ 
(B) & $\mu \geq 0$ \\ 
(C) & $\mathbf{\xi \geq 0}\smallskip $ \\ 
(D) & $\sum_{i=1}^{N}\xi _{i}=1\smallskip $ \\ 
(E) & $(\mathbf{x,\xi )\in \tciFourier }$%
\end{tabular}%
\end{array}
\label{maxMaxVersion}
\end{equation}

Two maximization operators can be collapsed into a single operator. 
\begin{equation}
\begin{array}{c}
\max\limits_{\mathbf{\xi ,x,}\mu ,\nu ,\mathbf{d}}-\mu
\sum_{i=1}^{N}q_{i}e^{-1-\frac{d_{i}+\nu }{\mu }}-\mu \Omega -\nu \\ 
\text{such that}\smallskip \\ 
\begin{tabular}{ll}
(A) & $d_{i}=z_{i}\exp \left[ -\alpha \sum\nolimits_{j=1}^{J}x_{j}\left(
\left( \mathbf{A}^{T}\mathbf{\xi }\right) _{j}-\mathbf{A}_{ij}\right) \right]
\text{ for }\forall i\in \left\{ 1,..,N\right\} $ \\ 
(B) & $\mu \geq 0$ \\ 
(C) & $\mathbf{\xi \geq 0}\smallskip $ \\ 
(D) & $\sum_{i=1}^{N}\xi _{i}=1\smallskip $ \\ 
(E) & $(\mathbf{x,\xi )\in \tciFourier }$%
\end{tabular}%
\end{array}
\label{maxMaxVersion2}
\end{equation}

\subsubsection{Further Simplification through Algebraic Manipulation}

Because the objective function of (\ref{maxMaxVersion2}) is a strictly
concave function of $\nu $, we can find the global optimizing value of $\nu $
from the first-order condition.%
\begin{equation*}
-\mu \sum_{i=1}^{N}q_{i}\left( -\frac{1}{\mu }\right) e^{-1-\frac{d_{i}+\nu 
}{\mu }}-1=0
\end{equation*}%
\begin{equation*}
\sum_{i=1}^{N}q_{i}e^{-1-\frac{d_{i}+\nu }{\mu }}=1
\end{equation*}%
\begin{equation*}
e^{-\frac{\nu }{\mu }}e^{-1}\sum_{i=1}^{N}q_{i}e^{-\frac{d_{i}}{\mu }}=1
\end{equation*}%
\begin{equation*}
-\frac{\nu }{\mu }-1+\ln \left( \sum_{i=1}^{N}q_{i}e^{-\frac{d_{i}}{\mu }%
}\right) =0
\end{equation*}%
\begin{equation}
\nu ^{\ast }=-\mu +\mu \ln \left( \sum_{i=1}^{N}q_{i}e^{-\frac{d_{i}}{\mu }%
}\right)  \label{optimalV}
\end{equation}

We substitute (\ref{optimalV}) into the objective function of (\ref%
{maxMaxVersion2}).%
\begin{eqnarray}
&&-\mu \sum_{i=1}^{N}q_{i}e^{-1-\frac{d_{i}+\nu ^{\ast }}{\mu }}-\mu \Omega
-\nu ^{\ast }  \notag \\
&=&-\mu \frac{e}{\sum_{i=1}^{N}q_{i}e^{-\frac{d_{i}}{\mu }}}%
\sum_{i=1}^{N}q_{i}e^{-1-\frac{d_{i}}{\mu }}-\mu \Omega +\mu -\mu \ln \left(
\sum_{i=1}^{N}q_{i}e^{-\frac{d_{i}}{\mu }}\right)  \notag \\
&=&-\mu -\mu \Omega +\mu -\mu \ln \left( \sum_{i=1}^{N}q_{i}e^{-\frac{d_{i}}{%
\mu }}\right)  \notag \\
&=&-\mu \Omega -\mu \ln \left( \sum_{i=1}^{N}q_{i}e^{-\frac{d_{i}}{\mu }%
}\right)  \label{gettingRidOfV}
\end{eqnarray}

Substituting (\ref{gettingRidOfV}) into (\ref{maxMaxVersion2}) further
simplifies the problem.

\begin{equation}
\begin{array}{c}
\min\limits_{\mathbf{\xi ,x,}\mu ,\mathbf{d}}\mu \Omega +\mu \ln \left(
\sum_{i=1}^{N}q_{i}e^{-\frac{d_{i}}{\mu }}\right) \\ 
\text{such that}\smallskip \\ 
\begin{tabular}{ll}
(A) & $d_{i}=z_{i}\exp \left[ -\alpha \sum\nolimits_{j=1}^{J}x_{j}\left(
\left( \mathbf{A}^{T}\mathbf{\xi }\right) _{j}-\mathbf{A}_{ij}\right) \right]
\text{ for }\forall i\in \left\{ 1,..,N\right\} $ \\ 
(B) & $\mu \geq 0$ \\ 
(C) & $\mathbf{\xi \geq 0}\smallskip $ \\ 
(D) & $\sum_{i=1}^{N}\xi _{i}=1\smallskip $ \\ 
(E) & $(\mathbf{x,\xi )\in \tciFourier }$%
\end{tabular}%
\end{array}
\label{collapsedMaxMaxVersion4}
\end{equation}

We define new constants.

\begin{equation}
\theta _{i}=e^{\Omega }q_{i}>0\text{ for }\forall i  \label{defineThetea}
\end{equation}

Then, the objective function of (\ref{collapsedMaxMaxVersion4}) can be more
succinctly represented as a function of $\mu $%
\begin{eqnarray}
\ell \left( \mu \right) &=&\mu \Omega +\mu \ln \left( \sum_{i=1}^{N}q_{i}e^{-%
\frac{d_{i}}{\mu }}\right) =\mu \ln \left( \sum_{i=1}^{N}q_{i}e^{\Omega }e^{-%
\frac{d_{i}}{\mu }}\right)  \notag \\
&=&\mu \ln \left( \sum_{i=1}^{N}\theta _{i}e^{-\frac{d_{i}}{\mu }}\right)
\label{defineLu}
\end{eqnarray}

The optimization problem then becomes:%
\begin{equation}
\begin{array}{c}
\min\limits_{\mathbf{\xi ,x,}\mu ,\mathbf{d}}\mu \ln \left(
\sum_{i=1}^{N}\theta _{i}e^{-\frac{d_{i}}{\mu }}\right) \\ 
\text{such that}\smallskip \\ 
\begin{tabular}{ll}
(A) & $d_{i}=z_{i}\exp \left[ -\alpha \sum\nolimits_{j=1}^{J}x_{j}\left(
\left( \mathbf{A}^{T}\mathbf{\xi }\right) _{j}-\mathbf{A}_{ij}\right) \right]
\text{ for }\forall i\in \left\{ 1,..,N\right\} $ \\ 
(B) & $\mu \geq 0$ \\ 
(C) & $\mathbf{\xi \geq 0}\smallskip $ \\ 
(D) & $\sum_{i=1}^{N}\xi _{i}=1\smallskip $ \\ 
(E) & $(\mathbf{x,\xi )\in \tciFourier }$%
\end{tabular}%
\end{array}
\label{collapsedMaxMaxVer5}
\end{equation}

\subsubsection{Linearization of Constraint (A)}

Note that constraint (A) in (\ref{collapsedMaxMaxVer5}) involves the
quadratic terms $\sum\nolimits_{j=1}^{J}x_{j}\left( \mathbf{A}^{T}\mathbf{%
\xi }\right) _{j}$. Therefore, in this section, we suggest a way to
linearize this constraint.

Intuitively, $x_{j}\left( \mathbf{A}^{T}\mathbf{\xi }\right) _{j}$ is simply
the market maker's revenue from the $j$th order. Let $R_{j}$ denote the
market maker's revenue from the $j$th order. $x_{j}$ is the number of shares
of the option traded. $\left( \mathbf{A}^{T}\mathbf{\xi }\right) _{j}$ is
the market-clearing price of the $j$th order.

To begin the transformation, we first consider the feasible set of $\mathbf{x%
}$ and $\mathbf{\xi }$. We say that the pair $\mathbf{x}$ and $\mathbf{\xi }$
are feasible if and only if the pair satisfies (\ref{constraintE}).

\begin{equation}
\begin{tabular}{ll}
(E1) & $\forall j\in \{1,2,..,J\},%
\begin{array}{ccc}
x_{j}=0 & \text{ if } & \left( \mathbf{A}^{T}\mathbf{\xi }\right)
_{j}>B_{j}b_{j}%
\end{array}%
$ \\ 
(E2) & $\forall j\in \{1,2,..,J\},%
\begin{array}{ccc}
x_{j}\in \left[ 0,Q_{j}\right] & \text{ if } & \left( \mathbf{A}^{T}\mathbf{%
\xi }\right) _{j}=B_{j}b_{j}%
\end{array}%
$ \\ 
(E3) & $\forall j\in \{1,2,..,J\},%
\begin{array}{ccc}
x_{j}=Q_{j} & \text{ if } & \left( \mathbf{A}^{T}\mathbf{\xi }\right)
_{j}<B_{j}b_{j}%
\end{array}%
$%
\end{tabular}
\label{constraintE}
\end{equation}

We restrict attention to the $j$th order. Figure 4 shows the feasible set of
pairs $x_{j}$ and $\left( \mathbf{A}^{T}\mathbf{\xi }\right) _{j}$. The
figure is just a graphic illustration of (\ref{constraintE}).\footnote{%
I can simply ignore cases for which $Q_{j}=0$. I can also simply remove the $%
j$th order from my optimization problem.}

\FRAME{ftbpFU}{4.2108in}{2.7778in}{0pt}{\Qcb{The feasible set of $x_{j}$ and 
$(A^{T}\protect\xi )_{j}$}}{}{feasibleset.bmp}{\special{language "Scientific
Word";type "GRAPHIC";maintain-aspect-ratio TRUE;display "USEDEF";valid_file
"F";width 4.2108in;height 2.7778in;depth 0pt;original-width
13.9442in;original-height 9.167in;cropleft "0";croptop "1";cropright
"1";cropbottom "0";filename 'feasibleSet.bmp';file-properties "XNPEU";}}

Figure 5 is a three-dimensional graph. The graph shows the market maker's
revenue as a function of the quantity filled ($x_{j}$) and the
market-clearing price $\left( \mathbf{A}^{T}\mathbf{\xi }\right) _{j}$.

\FRAME{ftbpFU}{4.625in}{3.1194in}{0pt}{\Qcb{The market maker's revenue from
the $j$th order as a function of the quantity filled $x_{j}$ and the
market-clearing price of the $j$th order $\left( \mathbf{\bar{A}}^{T}\mathbf{%
\protect\xi }\right) _{j}$}}{}{revenue3dgraph.bmp}{\special{language
"Scientific Word";type "GRAPHIC";maintain-aspect-ratio TRUE;display
"USEDEF";valid_file "F";width 4.625in;height 3.1194in;depth
0pt;original-width 13.8197in;original-height 9.2915in;cropleft "0";croptop
"1";cropright "1";cropbottom "0";filename
'revenue3Dgraph.bmp';file-properties "XNPEU";}}

\FRAME{ftbpFU}{4.4313in}{3.288in}{0pt}{\Qcb{The market maker's revenue from
the $j$th order as a function of the quantity filled $x_{j}$ and the
market-clearing price of the $j$th order $\left( \mathbf{A}^{T}\mathbf{%
\protect\xi }\right) _{j}$}}{}{revenue3dgraph2_revised.bmp}{\special%
{language "Scientific Word";type "GRAPHIC";maintain-aspect-ratio
TRUE;display "USEDEF";valid_file "F";width 4.4313in;height 3.288in;depth
0pt;original-width 13.7358in;original-height 10.1667in;cropleft "0";croptop
"1";cropright "1";cropbottom "0";filename
'revenue3Dgraph2_revised.bmp';file-properties "XNPEU";}}

For illustrative purposes, Figure 6 dissects Figure 5 into three distinct
regions. Region (1) corresponds to constraint (E1) in (\ref{constraintE}).
Region (2) and (3) correspond to constraints (E2) and (E3), respectively.

If we assume that the market-clearing price $\left( \mathbf{A}^{T}\mathbf{%
\xi }\right) _{j}$ and the quantity traded $x_{j}$ can take any real values,
the market maker's revenue $R_{j}$ becomes a nonlinear term $\left( \mathbf{A%
}^{T}\mathbf{\xi }\right) _{j}x_{j}$. However, if we restrict attention to
the feasible set in Figure 4, either the market-clearing price or the
quantity filled is held constant in each of the three regions. The market
maker's revenue $R_{j}$ is a piece-wise linear function.%
\begin{equation}
R_{j}=\left\{ 
\begin{array}{ccc}
0 & if & \left( \mathbf{A}^{T}\mathbf{\xi }\right) _{j}>B_{j}b_{j} \\ 
B_{j}b_{j}x_{j} & if & \left( \mathbf{A}^{T}\mathbf{\xi }\right)
_{j}=B_{j}b_{j} \\ 
Q_{j}\left( \mathbf{A}^{T}\mathbf{\xi }\right) _{j} & if & \left( \mathbf{A}%
^{T}\mathbf{\xi }\right) _{j}<B_{j}b_{j}%
\end{array}%
\right\}  \label{revenueFunction}
\end{equation}

(\ref{revenueFunction}) is equivalent to (\ref{revenueFunction2}) as long as 
$x_{j}$ and $\left( \mathbf{A}^{T}\mathbf{\xi }\right) _{j}$ belong to the
feasible set that Figure 4 represents. $\left[ \left( \mathbf{A}^{T}\mathbf{%
\xi }\right) _{j}-B_{j}b_{j}\right] ^{+}$ is a short-hand notation for $\max
\left\{ 0,\left( \mathbf{A}^{T}\mathbf{\xi }\right) _{j}-B_{j}b_{j}\right\} $%
. 
\begin{equation}
R_{j}=Q_{j}\left( \mathbf{A}^{T}\mathbf{\xi }\right) _{j}+B_{j}b_{j}\left(
x_{j}-Q_{j}\right) -Q_{j}\left[ \left( \mathbf{A}^{T}\mathbf{\xi }\right)
_{j}-B_{j}b_{j}\right] ^{+}  \label{revenueFunction2}
\end{equation}

For example, consider region (1) where $\left( \mathbf{A}^{T}\mathbf{\xi }%
\right) _{j}>B_{j}b_{j}$ and $x_{j}=0$. Then, (\ref{revenueFunction2})
reduces to (\ref{Region1Example}). Note that (\ref{Region1Example}) agrees
with (\ref{revenueFunction}).%
\begin{eqnarray}
R_{j} &=&Q_{j}\left( \mathbf{A}^{T}\mathbf{\xi }\right)
_{j}+B_{j}b_{j}\left( x_{j}-Q_{j}\right) -Q_{j}\left[ \left( \mathbf{A}^{T}%
\mathbf{\xi }\right) _{j}-B_{j}b_{j}\right] ^{+}  \notag \\
&=&Q_{j}\left( \mathbf{A}^{T}\mathbf{\xi }\right) _{j}+B_{j}b_{j}\left(
x_{j}-Q_{j}\right) -Q_{j}\left[ \left( \mathbf{A}^{T}\mathbf{\xi }\right)
_{j}-B_{j}b_{j}\right]  \notag \\
&=&B_{j}b_{j}x_{j}=B_{j}b_{j}\cdot 0=0  \label{Region1Example}
\end{eqnarray}

We can simplify (\ref{revenueFunction2}):%
\begin{eqnarray}
R_{j} &=&Q_{j}\left( \mathbf{A}^{T}\mathbf{\xi }\right)
_{j}+B_{j}b_{j}\left( x_{j}-Q_{j}\right) -Q_{j}\left[ \left( \mathbf{A}^{T}%
\mathbf{\xi }\right) _{j}-B_{j}b_{j}\right] ^{+}  \notag \\
&=&\min \left[ Q_{j}\left( \mathbf{A}^{T}\mathbf{\xi }\right)
_{j}+B_{j}b_{j}\left( x_{j}-Q_{j}\right) ,Q_{j}\left( \mathbf{A}^{T}\mathbf{%
\xi }\right) _{j}+B_{j}b_{j}\left( x_{j}-Q_{j}\right) -Q_{j}\left\{ \left( 
\mathbf{A}^{T}\mathbf{\xi }\right) _{j}-B_{j}b_{j}\right\} \right]  \notag \\
&=&\min \left[ Q_{j}\left( \mathbf{A}^{T}\mathbf{\xi }\right)
_{j}+B_{j}b_{j}\left( x_{j}-Q_{j}\right) ,B_{j}b_{j}x_{j}\right]
\label{RjInDifferentFormat}
\end{eqnarray}

Substitution of (\ref{RjInDifferentFormat}) into (\ref{collapsedMaxMaxVer5})
yields (\ref{collapsedMaxMaxVer6}). Constraint (E) in (\ref%
{collapsedMaxMaxVer6}) ensures that the pair $(\mathbf{x,\xi )}$ is within
the feasible set shown in Figure 1. Replacement of the quadratic term with
the piece-wise linear term is valid due to this restriction. 
\begin{equation}
\begin{array}{c}
\min\limits_{\mathbf{\xi ,x,}\mu ,\mathbf{d}}\mu \ln \left(
\sum_{i=1}^{N}\theta _{i}e^{-\frac{d_{i}}{\mu }}\right) \\ 
\text{such that}\smallskip \\ 
\begin{tabular}{ll}
(A) & $-d_{i}=-z_{i}e^{\alpha \sum\nolimits_{j=1}^{J}\left[ x_{j}\mathbf{A}%
_{ij}-\min \left\{ Q_{j}\left( \mathbf{A}^{T}\mathbf{\xi }\right)
_{j}+B_{j}b_{j}\left( x_{j}-Q_{j}\right) ,B_{j}b_{j}x_{j}\right\} \right] }%
\text{ for }\forall i$ \\ 
(B) & $\mu \geq 0$ \\ 
(C) & $\mathbf{\xi \geq 0}\smallskip $ \\ 
(D) & $\sum_{i=1}^{N}\xi _{i}=1\smallskip $ \\ 
(E) & $(\mathbf{x,\xi )\in \tciFourier }$%
\end{tabular}%
\end{array}
\label{collapsedMaxMaxVer6}
\end{equation}

Because $z_{i}$ is negative and $\alpha $ is positive, constraint (A) of (%
\ref{collapsedMaxMaxVer6}) is equivalent to (\ref{constraintA_alternative}).%
\begin{equation}
-d_{i}=\max \left[ -z_{i}e^{\alpha \sum\nolimits_{j=1}^{J}\left[ x_{j}%
\mathbf{A}_{ij}-Q_{j}\left( \mathbf{A}^{T}\mathbf{\xi }\right)
_{j}-B_{j}b_{j}\left( x_{j}-Q_{j}\right) \right] },-z_{i}e^{\alpha
\sum\nolimits_{j=1}^{J}\left[ x_{j}\mathbf{A}_{ij}-B_{j}b_{j}x_{j}\right] }%
\right]  \label{constraintA_alternative}
\end{equation}

To minimize the objective function of (\ref{collapsedMaxMaxVer6}), $-d_{i}$
must be minimized. Hence, the optimization problem can be further reduced to
(\ref{collapsedMaxMaxVer7}). Note that constraints (B) to (F) are all linear.%
\begin{equation}
\begin{array}{c}
\min\limits_{\mathbf{\xi ,x,}\mu ,\mathbf{d,\zeta }}\mu \ln \left(
\sum_{i=1}^{N}\theta _{i}e^{-\frac{d_{i}}{\mu }}\right) \\ 
\text{such that}\smallskip \\ 
\begin{tabular}{ll}
(A) & $-d_{i}=-z_{i}e^{\zeta _{i}}\text{ for }\forall i$ \\ 
(B) & $\zeta _{i}\geq \alpha \sum\nolimits_{j=1}^{J}\left[ x_{j}\mathbf{A}%
_{ij}-Q_{j}\left( \mathbf{A}^{T}\mathbf{\xi }\right) _{j}-B_{j}b_{j}\left(
x_{j}-Q_{j}\right) \right] $ for $\forall i$ \\ 
(C) & $\zeta _{i}\geq $ $\alpha \sum\nolimits_{j=1}^{J}\left[ x_{j}\mathbf{A}%
_{ij}-B_{j}b_{j}x_{j}\right] $ for $\forall i$ \\ 
(F) & $\mu \geq 0$ \\ 
(G) & $\mathbf{\xi \geq 0}\smallskip $ \\ 
(H) & $\sum_{i=1}^{N}\xi _{i}=1\smallskip $ \\ 
(I) & $(\mathbf{x,\xi )\in \tciFourier }$%
\end{tabular}%
\end{array}
\label{collapsedMaxMaxVer7}
\end{equation}

$\mathbf{\zeta }\in 
\mathbb{R}
^{N\times 1}$ is a dummy varible. 
\begin{equation*}
\mathbf{\zeta =}\left[ 
\begin{array}{ccc}
\zeta _{1} & ... & \zeta _{N}%
\end{array}%
\right]
\end{equation*}

\subsection{The Proof of Lemma 2}

In optimization problem (\ref{mainOptimizationVer2}), to minimize the
objective, $\left( -d_{i}\right) $ needs to be mininized. Define a new
vector $\mathbf{\omega \in 
\mathbb{R}
}^{N\times 1}$ such that $\mathbf{\omega =[}\omega _{1},...,\omega _{N}]$
Hence, (\ref{mainOptimizationVer2}) is equivalent to (\ref%
{collapsedMaxMaxVer8}). 
\begin{equation}
\begin{array}{c}
\min\limits_{\mathbf{\xi ,x,}\mu \mathbf{,\zeta ,\omega }}\mu \ln \left(
\sum_{i=1}^{N}\theta _{i}e^{\frac{\omega _{i}}{\mu }}\right) \\ 
\text{such that}\smallskip \\ 
\begin{tabular}{ll}
(A) & $\omega _{i}\geq -z_{i}e^{\zeta _{i}}\text{ for }\forall i$ \\ 
(B) & $\zeta _{i}\geq \alpha \sum\nolimits_{j=1}^{J}\left[ x_{j}\mathbf{A}%
_{ij}-Q_{j}\left( \mathbf{A}^{T}\mathbf{\xi }\right) _{j}-B_{j}b_{j}\left(
x_{j}-Q_{j}\right) \right] $, $\forall i$ \\ 
(C) & $\zeta _{i}\geq \alpha \sum\nolimits_{j=1}^{J}\left[ x_{j}\mathbf{A}%
_{ij}-B_{j}b_{j}x_{j}\right] $, $\forall i$ \\ 
(D) & $\mu \geq 0$ \\ 
(E) & $(\mathbf{x,\xi )\in C}$%
\end{tabular}%
\end{array}
\label{collapsedMaxMaxVer8}
\end{equation}%
My goal is to show that (\ref{collapsedMaxMaxVer8}) is a convex optimization
problem. A necessary preliminary step is to show that the set of pairs of $%
\omega _{i}$ and $\zeta _{i}$ that satisfy constraint (A) in (\ref%
{collapsedMaxMaxVer8}) constitute a convex set.

\begin{lemma}
The set of pairs of $\omega _{i}$ and $\zeta _{i}$ that satisfy constraint
(A) in (\ref{collapsedMaxMaxVer8}) form a convex set.

\begin{proof}
Define a new function. 
\begin{equation}
F(\omega _{i},\zeta _{i})=-\omega _{i}-z_{i}e^{\zeta _{i}}
\label{defineFunctionF}
\end{equation}%
To prove the lemma, it suffices to show that function $F$ is convex. The
Hessian is:%
\begin{equation}
\nabla ^{2}F=\left[ 
\begin{array}{cc}
\frac{\partial ^{2}F}{\partial \omega _{i}^{2}} & \frac{\partial ^{2}F}{%
\partial \omega _{i}\partial \zeta _{i}} \\ 
\frac{\partial ^{2}F}{\partial \omega _{i}\partial \zeta _{i}} & \frac{%
\partial ^{2}F}{\partial \zeta _{i}^{2}}%
\end{array}%
\right] =\left[ 
\begin{array}{cc}
0 & 0 \\ 
0 & -z_{i}e^{\zeta _{i}}%
\end{array}%
\right]  \label{hessianF}
\end{equation}%
Because $z_{i}$ is negative, $\nabla ^{2}F$ is positive semidefinite.
Therefore, $F$ is convex.
\end{proof}
\end{lemma}

The next step is to show that the objective function $\mu \ln \left(
\sum_{i=1}^{N}\theta _{i}e^{\frac{\omega _{i}}{\mu }}\right) $ is convex.
Define a new function. 
\begin{equation}
G\left( \omega _{1},\omega _{2},...,\omega _{N},\mu \right) =\mu \ln \left(
\sum_{i=1}^{N}\theta _{i}e^{\frac{\omega _{i}}{\mu }}\right)
\label{defineFunctionG}
\end{equation}

Before proceeding with the proof, we present a very useful result from Boyd
and Vandenberghe (2004).\footnote{%
See equation (4.44) on page 162}.

\begin{lemma}
Let the function $f(\mathbf{y})$ be defined as (\ref{defineFunctionSmallF})
where $\mathbf{a}_{k}\mathbf{,y}\in 
\mathbb{R}
^{n}$,$b_{k}\in 
\mathbb{R}
$.%
\begin{equation}
f(\mathbf{y})=\ln \left( \sum_{k=1}^{K}e^{\mathbf{a}_{k}^{T}\mathbf{y+}%
b_{k}}\right)  \label{defineFunctionSmallF}
\end{equation}%
$f(\mathbf{y})$ is a convex function.
\end{lemma}

Therefore, the Hessian of $f(\mathbf{y})$ must be positive semidefinite.

\begin{lemma}
Function $G$, which is defined as (\ref{defineFunctionG}), is convex.

\begin{proof}
Note that function $G$ can be expressed as (\ref{functionGAlternativeExp}).
With $\mu $ fixed, the structure of $G$ as a function of $\omega _{1},\omega
_{2},...,\omega _{N}$ is exactly analogous to (\ref{defineFunctionSmallF}). 
\begin{equation}
G\left( \omega _{1},\omega _{2},...,\omega _{N},\mu \right) =\mu \ln \left(
\sum_{i=1}^{N}e^{\frac{1}{\mu }\omega _{i}+\ln \theta _{i}}\right)
\label{functionGAlternativeExp}
\end{equation}%
\newline
With $\mu $ fixed, $G$ is a convex function of $\omega _{1},\omega
_{2},...,\omega _{N}$. Hence all the principal minors of the matrix in (\ref%
{positivePrincipalMinor}) are nonnegative$.$%
\begin{equation}
\left[ 
\begin{array}{cccc}
\frac{\partial ^{2}G}{\partial \omega _{1}^{2}} & \frac{\partial ^{2}G}{%
\partial \omega _{1}\partial \omega _{2}} & ... & \frac{\partial ^{2}G}{%
\partial \omega _{1}\partial \omega _{N}} \\ 
\frac{\partial ^{2}G}{\partial \omega _{2}\partial \omega _{1}} & \frac{%
\partial ^{2}G}{\partial \omega _{2}^{2}} & ... & \frac{\partial ^{2}G}{%
\partial \omega _{2}\partial \omega _{N}} \\ 
... &  &  &  \\ 
\frac{\partial ^{2}G}{\partial \omega _{N}\partial \omega _{1}} & \frac{%
\partial ^{2}G}{\partial \omega _{N}\partial \omega _{2}} & ... & \frac{%
\partial ^{2}G}{\partial \omega _{N}^{2}}%
\end{array}%
\right]  \label{positivePrincipalMinor}
\end{equation}%
To prove the lemma, it suffices to show that the Hessian $\nabla ^{2}G$ is
positive semidefinite. We need to show that all the principal minors of $%
\nabla ^{2}G$ in (\ref{hessianG}) are nonnegative. 
\begin{equation}
\nabla ^{2}G=%
\begin{array}{cc}
& 
\begin{array}{ccc}
\omega _{1} & \text{ \ \ \ \ \ \ \ \ \ \ \ \ }\omega _{N} & \text{ \ \ \ \ }%
\mu%
\end{array}
\\ 
\begin{array}{c}
\omega _{1} \\ 
... \\ 
\omega _{N} \\ 
\mu%
\end{array}
& \left[ 
\begin{array}{cccc}
\frac{\partial ^{2}G}{\partial \omega _{1}^{2}} & ... & \frac{\partial ^{2}G%
}{\partial \omega _{1}\partial \omega _{N}} & \frac{\partial ^{2}G}{\partial
\omega _{1}\partial \mu } \\ 
... &  &  & ... \\ 
\frac{\partial ^{2}G}{\partial \omega _{N}\partial \omega _{1}} & ... & 
\frac{\partial ^{2}G}{\partial \omega _{N}^{2}} & \frac{\partial ^{2}G}{%
\partial \omega _{N}\partial \mu } \\ 
\frac{\partial ^{2}G}{\partial \mu \partial \omega _{1}} & ... & \frac{%
\partial ^{2}G}{\partial \mu \partial \omega _{N}} & \frac{\partial ^{2}G}{%
\partial \mu ^{2}}%
\end{array}%
\right]%
\end{array}
\label{hessianG}
\end{equation}%
However, because we already know that all the principal minors of the matrix
in (\ref{positivePrincipalMinor}) are nonnegative, it only remains to show
that $\frac{\partial ^{2}G}{\partial \mu ^{2}}\geq 0$ and $\det \nabla
^{2}G\geq 0$. \newline
\textbf{First, we show that }$\frac{\partial ^{2}G}{\partial \mu ^{2}}\geq 0$%
\newline
Find the first derivative of $G.$%
\begin{eqnarray}
\frac{\partial G}{\partial \mu } &=&\ln \left( \sum_{i=1}^{N}\theta _{i}e^{%
\frac{\omega _{i}}{\mu }}\right) +\mu \frac{\sum_{i=1}^{N}\theta _{i}\frac{%
-\omega _{i}}{\mu ^{2}}e^{\frac{\omega _{i}}{\mu }}}{\sum_{i=1}^{N}\theta
_{i}e^{\frac{\omega _{i}}{\mu }}}  \notag \\
&=&\ln \left( \sum_{i=1}^{N}\theta _{i}e^{\frac{\omega _{i}}{\mu }}\right) -%
\frac{1}{\mu }\frac{\sum_{i=1}^{N}\theta _{i}\omega _{i}e^{\frac{\omega _{i}%
}{\mu }}}{\sum_{i=1}^{N}\theta _{i}e^{\frac{\omega _{i}}{\mu }}}
\label{dGdMu}
\end{eqnarray}%
\newline
Find the second derivative of $G$.%
\begin{eqnarray}
\frac{\partial ^{2}G}{\partial \mu ^{2}} &=&\frac{\sum_{i=1}^{N}\theta _{i}%
\frac{-\omega _{i}}{\mu ^{2}}e^{\frac{\omega _{i}}{\mu }}}{%
\sum_{i=1}^{N}\theta _{i}e^{\frac{\omega _{i}}{\mu }}}+\frac{1}{\mu ^{2}}%
\frac{\sum_{i=1}^{N}\theta _{i}\omega _{i}e^{\frac{\omega _{i}}{\mu }}}{%
\sum_{i=1}^{N}\theta _{i}e^{\frac{\omega _{i}}{\mu }}}  \notag \\
&&-\frac{1}{\mu }\frac{\sum_{i=1}^{N}\theta _{i}e^{\frac{\omega _{i}}{\mu }%
}\sum_{i=1}^{N}\theta _{i}\frac{-\omega _{i}^{2}}{\mu ^{2}}e^{\frac{\omega
_{i}}{\mu }}-\sum_{i=1}^{N}\theta _{i}\frac{-\omega _{i}}{\mu ^{2}}e^{\frac{%
\omega _{i}}{\mu }}\sum_{i=1}^{N}\theta _{i}\omega _{i}e^{\frac{\omega _{i}}{%
\mu }}}{\left[ \sum_{i=1}^{N}\theta _{i}e^{\frac{\omega _{i}}{\mu }}\right]
^{2}}  \label{LuFirstDerivative}
\end{eqnarray}%
\begin{eqnarray}
\frac{\partial ^{2}G}{\partial \mu ^{2}} &=&\frac{1}{\mu }\frac{%
\sum_{i=1}^{N}\theta _{i}e^{\frac{\omega _{i}}{\mu }}\sum_{i=1}^{N}\theta
_{i}\frac{\omega _{i}^{2}}{\mu ^{2}}e^{\frac{\omega _{i}}{\mu }%
}-\sum_{i=1}^{N}\theta _{i}\frac{\omega _{i}}{\mu ^{2}}e^{\frac{\omega _{i}}{%
\mu }}\sum_{i=1}^{N}\theta _{i}\omega _{i}e^{\frac{\omega _{i}}{\mu }}}{%
\left[ \sum_{i=1}^{N}\theta _{i}e^{\frac{\omega _{i}}{\mu }}\right] ^{2}} 
\notag \\
&=&\frac{\sum_{i=1}^{N}\theta _{i}e^{\frac{\omega _{i}}{\mu }%
}\sum_{i=1}^{N}\theta _{i}\omega _{i}^{2}e^{\frac{\omega _{i}}{\mu }}-\left[
\sum_{i=1}^{N}\theta _{i}\omega _{i}e^{\frac{\omega _{i}}{\mu }}\right] ^{2}%
}{\mu ^{3}\left[ \sum_{i=1}^{N}\theta _{i}e^{\frac{\omega _{i}}{\mu }}\right]
^{2}}  \label{LuSecondDerivative}
\end{eqnarray}%
The denominator of (\ref{LuSecondDerivative}) is positive. Hence, it only
remains to show that the numerator is nonnegative. This part can be shown by
using a Cauchy-Schwarz inequality.%
\begin{equation}
\sum_{i=1}^{N}\left( \sqrt{\theta _{i}e^{\frac{\omega _{i}}{\mu }}}\right)
^{2}\sum_{i=1}^{N}\left( \sqrt{\theta _{i}\omega _{i}^{2}e^{\frac{\omega _{i}%
}{\mu }}}\right) ^{2}\geq \left[ \sum_{i=1}^{N}\sqrt{\theta _{i}e^{\frac{%
\omega _{i}}{\mu }}}\cdot \sqrt{\theta _{i}\omega _{i}^{2}e^{\frac{\omega
_{i}}{\mu }}}\right] ^{2}  \label{CauchySchwarz}
\end{equation}%
$\therefore \frac{\partial ^{2}G}{\partial \mu ^{2}}$ is always nonnegative.%
\newline
\textbf{Second, we show that }$\det \nabla ^{2}G\geq 0$\textbf{. }\newline
From (\ref{dGdMu}), we calculate $\frac{\partial ^{2}G}{\partial \omega
_{k}\partial \mu }$ where $k\in \{1,2,...,N\}.$%
\begin{eqnarray}
\frac{\partial ^{2}G}{\partial \omega _{k}\partial \mu } &=&\frac{\partial }{%
\partial \omega _{k}}\left[ \ln \left( \sum_{i=1}^{N}\theta _{i}e^{\frac{%
\omega _{i}}{\mu }}\right) -\frac{1}{\mu }\frac{\sum_{i=1}^{N}\theta
_{i}\omega _{i}e^{\frac{\omega _{i}}{\mu }}}{\sum_{i=1}^{N}\theta _{i}e^{%
\frac{\omega _{i}}{\mu }}}\right]  \notag \\
&=&\frac{\theta _{k}\frac{1}{\mu }e^{\frac{\omega _{k}}{\mu }}}{%
\sum_{i=1}^{N}\theta _{i}e^{\frac{\omega _{i}}{\mu }}}-\frac{1}{\mu }\frac{%
\left( \sum_{i=1}^{N}\theta _{i}e^{\frac{\omega _{i}}{\mu }}\right) \cdot 
\frac{\partial }{\partial \omega _{k}}\theta _{k}\omega _{k}e^{\frac{\omega
_{k}}{\mu }}-\frac{\theta _{k}}{\mu }e^{\frac{\omega _{k}}{\mu }}\left(
\sum_{i=1}^{N}\theta _{i}\omega _{i}e^{\frac{\omega _{i}}{\mu }}\right) }{%
\left[ \sum_{i=1}^{N}\theta _{i}e^{\frac{\omega _{i}}{\mu }}\right] ^{2}} 
\notag \\
&=&\frac{\theta _{k}\frac{1}{\mu }e^{\frac{\omega _{k}}{\mu }}}{%
\sum_{i=1}^{N}\theta _{i}e^{\frac{\omega _{i}}{\mu }}}-\frac{1}{\mu }\frac{%
\left( \sum_{i=1}^{N}\theta _{i}e^{\frac{\omega _{i}}{\mu }}\right) \left(
\theta _{k}e^{\frac{\omega _{k}}{\mu }}+\frac{\theta _{k}\omega _{k}}{\mu }%
e^{\frac{\omega _{k}}{\mu }}\right) -\frac{\theta _{k}}{\mu }e^{\frac{\omega
_{k}}{\mu }}\left( \sum_{i=1}^{N}\theta _{i}\omega _{i}e^{\frac{\omega _{i}}{%
\mu }}\right) }{\left[ \sum_{i=1}^{N}\theta _{i}e^{\frac{\omega _{i}}{\mu }}%
\right] ^{2}}  \notag \\
&=&\frac{\theta _{k}\frac{1}{\mu }e^{\frac{\omega _{k}}{\mu }}}{%
\sum_{i=1}^{N}\theta _{i}e^{\frac{\omega _{i}}{\mu }}}-\frac{\theta _{k}e^{%
\frac{\omega _{k}}{\mu }}+\frac{\theta _{k}\omega _{k}}{\mu }e^{\frac{\omega
_{k}}{\mu }}}{\mu \sum_{i=1}^{N}\theta _{i}e^{\frac{\omega _{i}}{\mu }}}+%
\frac{\theta _{k}e^{\frac{\omega _{k}}{\mu }}\left( \sum_{i=1}^{N}\theta
_{i}\omega _{i}e^{\frac{\omega _{i}}{\mu }}\right) }{\mu ^{2}\left[
\sum_{i=1}^{N}\theta _{i}e^{\frac{\omega _{i}}{\mu }}\right] ^{2}}  \notag \\
&=&-\frac{\theta _{k}\omega _{k}e^{\frac{\omega _{k}}{\mu }%
}\sum_{i=1}^{N}\theta _{i}e^{\frac{\omega _{i}}{\mu }}}{\mu ^{2}\left[
\sum_{i=1}^{N}\theta _{i}e^{\frac{\omega _{i}}{\mu }}\right] ^{2}}+\frac{%
\theta _{k}e^{\frac{\omega _{k}}{\mu }}\left( \sum_{i=1}^{N}\theta
_{i}\omega _{i}e^{\frac{\omega _{i}}{\mu }}\right) }{\mu ^{2}\left[
\sum_{i=1}^{N}\theta _{i}e^{\frac{\omega _{i}}{\mu }}\right] ^{2}}  \notag \\
&=&\frac{\theta _{k}e^{\frac{\omega _{k}}{\mu }}\left( \sum_{i=1}^{N}\theta
_{i}\omega _{i}e^{\frac{\omega _{i}}{\mu }}-\omega _{k}\sum_{i=1}^{N}\theta
_{i}e^{\frac{\omega _{i}}{\mu }}\right) }{\mu ^{2}\left[ \sum_{i=1}^{N}%
\theta _{i}e^{\frac{\omega _{i}}{\mu }}\right] ^{2}}  \notag \\
&=&\frac{\theta _{k}e^{\frac{\omega _{k}}{\mu }}\sum_{i=1}^{N}\theta
_{i}\left( \omega _{i}-\omega _{k}\right) e^{\frac{\omega _{i}}{\mu }}}{\mu
^{2}\left[ \sum_{i=1}^{N}\theta _{i}e^{\frac{\omega _{i}}{\mu }}\right] ^{2}}
\end{eqnarray}%
Similarly,%
\begin{equation}
\frac{\partial G}{\partial \omega _{k}}=\frac{\partial }{\partial \omega _{k}%
}\mu \ln \left( \sum_{i=1}^{N}\theta _{i}e^{\frac{\omega _{i}}{\mu }}\right)
=\mu \frac{\theta _{k}\frac{1}{\mu }e^{\frac{\omega _{k}}{\mu }}}{%
\sum_{i=1}^{N}\bar{\theta}_{i}e^{\frac{\omega _{i}}{\mu }}}=\frac{\theta
_{k}e^{\frac{\omega _{k}}{\mu }}}{\sum_{i=1}^{N}\theta _{i}e^{\frac{\omega
_{i}}{\mu }}}  \notag
\end{equation}%
\begin{eqnarray}
\frac{\partial ^{2}G}{\partial \omega _{k}^{2}} &=&\frac{\partial }{\partial
\omega _{k}}\frac{\theta _{k}e^{\frac{\omega _{k}}{\mu }}}{\sum_{i=1}^{N}%
\bar{\theta}_{i}e^{\frac{\omega _{i}}{\mu }}}  \notag \\
&=&\frac{\sum_{i=1}^{N}\theta _{i}e^{\frac{\omega _{i}}{\mu }}\cdot \frac{%
\theta _{k}}{\mu }e^{\frac{\omega _{k}}{\mu }}-\theta _{k}e^{\frac{\omega
_{k}}{\mu }}\cdot \frac{\theta _{k}}{\mu }e^{\frac{\omega _{k}}{\mu }}}{%
\left[ \sum_{i=1}^{N}\theta _{i}e^{\frac{\omega _{i}}{\mu }}\right] ^{2}} 
\notag \\
&=&\theta _{k}e^{\frac{\omega _{k}}{\mu }}\frac{\sum_{i=1}^{N}\theta _{i}e^{%
\frac{\omega _{i}}{\mu }}-\theta _{k}e^{\frac{\omega _{k}}{\mu }}}{\mu \left[
\sum_{i=1}^{N}\theta _{i}e^{\frac{\omega _{i}}{\mu }}\right] ^{2}}  \notag
\end{eqnarray}%
\newline
Provided that $j,k\in \{1,...,N\}$ and $j\neq k$,%
\begin{eqnarray}
\frac{\partial ^{2}G}{\partial \omega _{j}\partial \omega _{k}} &=&\frac{%
\partial }{\partial \omega _{j}}\frac{\theta _{k}e^{\frac{\omega _{k}}{\mu }}%
}{\sum_{i=1}^{N}\theta _{i}e^{\frac{\omega _{i}}{\mu }}}=\frac{%
\sum_{i=1}^{N}\theta _{i}e^{\frac{\omega _{i}}{\mu }}\cdot 0-\theta _{k}e^{%
\frac{\omega _{k}}{\mu }}\frac{1}{\mu }\theta _{j}e^{\frac{\omega _{j}}{\mu }%
}}{\left[ \sum_{i=1}^{N}\theta _{i}e^{\frac{\omega _{i}}{\mu }}\right] ^{2}}
\notag \\
&=&-\frac{1}{\mu }\frac{\theta _{k}\theta _{j}}{\left[ \sum_{i=1}^{N}\theta
_{i}e^{\frac{\omega _{i}}{\mu }}\right] ^{2}}e^{\frac{\omega _{k}}{\mu }}e^{%
\frac{\omega _{j}}{\mu }}  \notag
\end{eqnarray}%
\newline
Thus, $\nabla ^{2}G_{\cdot ,k}$ in (\ref{kthColumnHessianG}) shows the $k$th
column of $\nabla ^{2}G$. $\nabla ^{2}G_{\cdot ,1}$ denotes the first
column, $\nabla ^{2}G_{\cdot ,2}$ denotes the second column, and so forth.%
\begin{eqnarray}
\nabla ^{2}G_{\cdot ,k} &=&\left[ 
\begin{array}{c}
\frac{\partial ^{2}G}{\partial \omega _{1}\partial \omega _{k}} \\ 
... \\ 
\frac{\partial ^{2}G}{\partial \omega _{k}^{2}} \\ 
... \\ 
\frac{\partial ^{2}G}{\partial \omega _{N}\partial \omega _{k}} \\ 
\frac{\partial ^{2}G}{\partial \mu \partial \omega _{k}}%
\end{array}%
\right] =\left[ 
\begin{array}{c}
-\frac{\theta _{1}\theta _{k}e^{\frac{\omega _{k}}{\mu }}e^{\frac{\omega _{1}%
}{\mu }}}{\mu \left\{ \sum_{i=1}^{N}\theta _{i}e^{\frac{\omega _{i}}{\mu }%
}\right\} ^{2}} \\ 
... \\ 
\theta _{k}e^{\frac{\omega _{k}}{\mu }}\frac{\sum_{i=1}^{N}\theta _{i}e^{%
\frac{\omega _{i}}{\mu }}-\theta _{k}e^{\frac{\omega _{k}}{\mu }}}{\mu
\left\{ \sum_{i=1}^{N}\theta _{i}e^{\frac{\omega _{i}}{\mu }}\right\} ^{2}}
\\ 
... \\ 
-\frac{\theta _{N}\theta _{k}e^{\frac{\omega _{k}}{\mu }}e^{\frac{\omega _{N}%
}{\mu }}}{\mu \left\{ \sum_{i=1}^{N}\theta _{i}e^{\frac{\omega _{i}}{\mu }%
}\right\} ^{2}} \\ 
\frac{\theta _{k}e^{\frac{\omega _{k}}{\mu }}\sum_{i=1}^{N}\theta _{i}\left(
\omega _{i}-\omega _{k}\right) e^{\frac{\omega _{i}}{\mu }}}{\mu ^{2}\left\{
\sum_{i=1}^{N}\theta _{i}e^{\frac{\omega _{i}}{\mu }}\right\} ^{2}}%
\end{array}%
\right]  \notag \\
&=&\frac{\theta _{k}e^{\frac{\omega _{k}}{\mu }}}{\mu \left\{
\sum_{i=1}^{N}\theta _{i}e^{\frac{\omega _{i}}{\mu }}\right\} ^{2}}\left[ 
\begin{array}{c}
-\theta _{1}e^{\frac{\omega _{1}}{\mu }} \\ 
... \\ 
\sum_{i=1}^{N}\theta _{i}e^{\frac{\omega _{i}}{\mu }}-\theta _{k}e^{\frac{%
\omega _{k}}{\mu }} \\ 
... \\ 
-\theta _{N}e^{\frac{\omega _{N}}{\mu }} \\ 
\frac{1}{\mu }\sum_{i=1}^{N}\theta _{i}\left( \omega _{i}-\omega _{k}\right)
e^{\frac{\omega _{i}}{\mu }}%
\end{array}%
\right]  \label{kthColumnHessianG}
\end{eqnarray}%
\newline
Consider the following linear combination of the columns.%
\begin{eqnarray}
\sum_{k=1}^{N}\frac{\omega _{k}}{\mu }\nabla ^{2}G_{\cdot ,k}
&=&\sum_{k=1}^{N}\frac{\omega _{k}}{\mu }\left[ 
\begin{array}{c}
\frac{\partial ^{2}G}{\partial \omega _{1}\partial \omega _{k}} \\ 
... \\ 
\frac{\partial ^{2}G}{\partial \omega _{k}^{2}} \\ 
... \\ 
\frac{\partial ^{2}G}{\partial \omega _{N}\partial \omega _{k}} \\ 
\frac{\partial ^{2}G}{\partial \mu \partial \omega _{k}}%
\end{array}%
\right]  \notag \\
&=&\sum_{k=1}^{N}\frac{\theta _{k}\omega _{k}e^{\frac{\omega _{k}}{\mu }}}{%
\mu ^{2}\left\{ \sum_{i=1}^{N}\theta _{i}e^{\frac{\omega _{i}}{\mu }%
}\right\} ^{2}}\left[ 
\begin{array}{c}
-\theta _{1}e^{\frac{\omega _{1}}{\mu }} \\ 
... \\ 
\sum_{i=1}^{N}\theta _{i}e^{\frac{\omega _{i}}{\mu }}-\theta _{k}e^{\frac{%
\omega _{k}}{\mu }} \\ 
... \\ 
-\theta _{N}e^{\frac{\omega _{N}}{\mu }} \\ 
\frac{1}{\mu }\sum_{i=1}^{N}\theta _{i}\left( \omega _{i}-\omega _{k}\right)
e^{\frac{\omega _{i}}{\mu }}%
\end{array}%
\right]  \notag \\
&=&\frac{1}{\mu ^{2}\left\{ \sum_{i=1}^{N}\theta _{i}e^{\frac{\omega _{i}}{%
\mu }}\right\} ^{2}}\left[ 
\begin{array}{c}
\theta _{1}\omega _{1}e^{\frac{\omega _{1}}{\mu }}\sum_{i=1}^{N}\theta
_{i}e^{\frac{\omega _{i}}{\mu }}-\theta _{1}e^{\frac{\omega _{1}}{\mu }%
}\sum_{k=1}^{N}\theta _{k}\omega _{k}e^{\frac{\omega _{k}}{\mu }} \\ 
... \\ 
\theta _{_{N}}\omega _{_{N}}e^{\frac{\omega _{_{N}}}{\mu }%
}\sum_{i=1}^{N}\theta _{i}e^{\frac{\omega _{i}}{\mu }}-\theta _{N}e^{\frac{%
\omega _{_{N}}}{\mu }}\sum_{k=1}^{N}\theta _{k}\omega _{k}e^{\frac{\omega
_{k}}{\mu }} \\ 
\frac{1}{\mu }\sum_{k=1}^{N}\left\{ \theta _{k}\omega _{k}e^{\frac{\omega
_{k}}{\mu }}\sum_{i=1}^{N}\theta _{i}\left( \omega _{i}-\omega _{k}\right)
e^{\frac{\omega _{i}}{\mu }}\right\}%
\end{array}%
\right]  \notag \\
&=&\frac{1}{\mu ^{2}\left\{ \sum_{i=1}^{N}\theta _{i}e^{\frac{\omega _{i}}{%
\mu }}\right\} ^{2}}\left[ 
\begin{array}{c}
\theta _{1}e^{\frac{\omega _{1}}{\mu }}\left\{ \sum_{i=1}^{N}\omega
_{1}\theta _{i}e^{\frac{\omega _{i}}{\mu }}-\sum_{k=1}^{N}\theta _{k}\omega
_{k}e^{\frac{\omega _{k}}{\mu }}\right\} \\ 
... \\ 
\theta _{_{N}}\omega _{_{N}}\left\{ \sum_{i=1}^{N}\omega _{N}\theta _{i}e^{%
\frac{\omega _{i}}{\mu }}-\sum_{k=1}^{N}\theta _{k}\omega _{k}e^{\frac{%
\omega _{k}}{\mu }}\right\} \\ 
\frac{1}{\mu }\sum_{k=1}^{N}\left\{ \theta _{k}\omega _{k}e^{\frac{\omega
_{k}}{\mu }}\left( \sum_{i=1}^{N}\theta _{i}\omega _{i}e^{\frac{\omega _{i}}{%
\mu }}-\omega _{k}\sum_{i=1}^{N}\theta _{i}e^{\frac{\omega _{i}}{\mu }%
}\right) \right\}%
\end{array}%
\right]  \notag
\end{eqnarray}%
\begin{eqnarray}
&&\sum_{k=1}^{N}\frac{\omega _{k}}{\mu }\nabla ^{2}G_{\cdot ,k}  \notag \\
&=&\frac{1}{\mu ^{2}\left\{ \sum_{i=1}^{N}\theta _{i}e^{\frac{\omega _{i}}{%
\mu }}\right\} ^{2}}\left[ 
\begin{array}{c}
\theta _{1}e^{\frac{\omega _{1}}{\mu }}\sum_{k=1}^{N}\theta _{k}\left(
\omega _{1}-\omega _{k}\right) e^{\frac{\omega _{k}}{\mu }} \\ 
... \\ 
\theta _{_{N}}\omega _{_{N}}\sum_{k=1}^{N}\theta _{k}\left( \omega
_{N}-\omega _{k}\right) e^{\frac{\omega _{k}}{\mu }} \\ 
\frac{1}{\mu }\left[ \left\{ \sum_{k=1}^{N}\theta _{k}\omega _{k}e^{\frac{%
\omega _{k}}{\mu }}\right\} ^{2}-\left\{ \sum_{k=1}^{N}\theta _{k}e^{\frac{%
\omega _{k}}{\mu }}\right\} \left\{ \sum_{k=1}^{N}\theta _{k}\omega
_{k}^{2}e^{\frac{\omega _{k}}{\mu }}\right\} \right]%
\end{array}%
\right]  \label{columnLinCombination}
\end{eqnarray}%
\newline
However, the last column of $\nabla ^{2}G$ is%
\begin{eqnarray}
\nabla ^{2}G_{\cdot N+1} &=&\left[ 
\begin{array}{c}
\frac{\partial ^{2}G}{\partial \omega _{1}\partial \mu } \\ 
\frac{\partial ^{2}G}{\partial \omega _{2}\partial \mu } \\ 
... \\ 
\frac{\partial ^{2}G}{\partial \omega _{N}\partial \mu } \\ 
\frac{\partial ^{2}G}{\partial \mu ^{2}}%
\end{array}%
\right]  \notag \\
&=&\frac{1}{\mu ^{2}\left[ \sum_{i=1}^{N}\theta _{i}e^{\frac{\omega _{i}}{%
\mu }}\right] ^{2}}\left[ 
\begin{array}{c}
\theta _{1}e^{\frac{\omega _{1}}{\mu }}\sum_{i=1}^{N}\theta _{i}\left(
\omega _{i}-\omega _{1}\right) e^{\frac{\omega _{i}}{\mu }} \\ 
\theta _{2}e^{\frac{\omega _{2}}{\mu }}\sum_{i=1}^{N}\theta _{i}\left(
\omega _{i}-\omega _{2}\right) e^{\frac{\omega _{i}}{\mu }} \\ 
... \\ 
\theta _{N}e^{\frac{\omega _{N}}{\mu }}\sum_{i=1}^{N}\theta _{i}\left(
\omega _{i}-\omega _{N}\right) e^{\frac{\omega _{i}}{\mu }} \\ 
\frac{1}{\mu }\sum_{i=1}^{N}\theta _{i}e^{\frac{\omega _{i}}{\mu }%
}\sum_{i=1}^{N}\theta _{i}\omega _{i}^{2}e^{\frac{\omega _{i}}{\mu }}-\frac{1%
}{\mu }\left[ \sum_{i=1}^{N}\theta _{i}\omega _{i}e^{\frac{\omega _{i}}{\mu }%
}\right] ^{2}%
\end{array}%
\right]  \label{lastColumnHessian}
\end{eqnarray}%
The combination of (\ref{columnLinCombination}) and (\ref{lastColumnHessian}%
) yields (\ref{linCombinationZeroColumn}).%
\begin{equation}
\sum_{k=1}^{N}\frac{\omega _{k}}{\mu }\nabla ^{2}G_{\cdot ,k}+\nabla
^{2}G_{\cdot N+1}=\left[ 
\begin{array}{c}
0 \\ 
0 \\ 
... \\ 
0 \\ 
0%
\end{array}%
\right]  \label{linCombinationZeroColumn}
\end{equation}%
Therefore,%
\begin{eqnarray}
\det \nabla ^{2}G &=&\det \left[ 
\begin{array}{ccccc}
\nabla ^{2}G_{\cdot ,1} & \nabla ^{2}G_{\cdot ,2} & ... & \nabla
^{2}G_{\cdot ,N} & \nabla ^{2}G_{\cdot ,N+1}%
\end{array}%
\right]  \notag \\
&=&\det \left[ 
\begin{array}{ccccc}
\nabla ^{2}G_{\cdot ,1} & \nabla ^{2}G_{\cdot ,2} & ... & \nabla
^{2}G_{\cdot ,N} & \sum_{k=1}^{N}\frac{\omega _{k}}{\mu }\nabla ^{2}G_{\cdot
,k}+\nabla ^{2}G_{\cdot N+1}%
\end{array}%
\right]  \notag \\
&=&\det \left[ 
\begin{array}{ccccc}
\nabla ^{2}G_{\cdot ,1} & \nabla ^{2}G_{\cdot ,2} & ... & \nabla
^{2}G_{\cdot ,N} & 0%
\end{array}%
\right]  \notag \\
&=&0  \label{zeroDeterminantHG}
\end{eqnarray}%
Because both $\frac{\partial ^{2}G}{\partial \mu ^{2}}$ and $\det \nabla
^{2}G$ are nonnegative, $G$ is a convex function.
\end{proof}
\end{lemma}

(\ref{mainOptimizationVer2}) is a convex optimization problem because both
the objective function and the feasible set are convex.

\subsection{The Proof of Theorem 3}

Note from our pseudo-code that the number of times we need to solve the
problem (\ref{pseudoCodeSubProblem}) is $\Pi _{k=1}^{K}n_{k}$%
\begin{equation}
\Pi _{k=1}^{K}n_{k}\leq \Pi _{k=1}^{K}J=J^{K}  \label{polynomialBound1}
\end{equation}

In addition, it is well known that, in principle, the interior-point method
can solve any convex optimization in polynomial time of the problem
dimension. Thus, (\ref{pseudoCodeSubProblem}) should also be solvable in
polynomial time. Let $\tau (\ell _{1},\ell _{2},...,\ell _{K})$ denote the
time required to solve problem (\ref{pseudoCodeSubProblem}).

Let $T$ denote the time required to execute the entire pseudo-code.%
\begin{equation*}
T=\sum\limits_{\ell _{1}=1}^{n_{1}}\sum\limits_{\ell
_{2}=1}^{n_{2}}...\sum\limits_{\ell _{K}=1}^{n_{K}}\tau (\ell _{1},\ell
_{2},...\ell _{K})\leq J^{K}\max\limits_{\ell _{1},\ell _{2},...,\ell
_{K}}\tau (\ell _{1},\ell _{2},...\ell _{K})
\end{equation*}

$\max\limits_{\ell _{1},\ell _{2},...,\ell _{K}}\tau (\ell _{1},\ell
_{2},...\ell _{K})$ is bounded above by a polynomial function of $J$. Thus, $%
T$ is also bounded above by a polynomial of $J$.

\subsection{Other Well-Known Strengths of the Pari-mutuel Auction}

Please see Baron and Lange (2007) or Lange and Economide (2005) for a more
thorough discussion. In this subsection, we briefly introduce some of the
strengths of pari-mutuel auctions and our insights.

\subsubsection{Liquidity Aggregation}

The market maker can reduce his/her inventory holding cost by being involved
in more than one market. This lower inventory holding cost allows the market
maker to supply liquidity to each market at lower cost.

To illustrate, consider an exotic derivative market with the Consumer Price
Index (CPI) as the underlying variable. Suppose that there are two types of
options: a call option with the strike 0\% and a put option with the same
strike. For example, if the CPI is 1\%, the call option pays \$1, while the
put option does not pay. Imagine that there is an overwhelming demand for
both options.

First, consider the case in which two markets are fragmented. There is one
dealer for each market. Overwhelming demand for each option forces the
market maker to take a large short position. The inventory of each market
maker becomes highly unbalanced, exposing him/her to significant risk. This
increased inventory cost leads to a larger bid-ask spread and reduced
liquidity in each market (Stoll, 1978).

In contrast, consider having a common market maker serve both markets.
Simultaneously taking large short positions in both the call option and put
options is less risky than shorting only one option. As the underlying
variable fluctuates, the price of the call and that of the put move in the
opposite direction. Therefore, holding the call option can partly offset the
risk of holding the put option and vice versa. A smaller inventory holding
cost leads to a narrower bid-ask spread and enhanced liquidity in each
market.

This effect is called "liquidity aggregation" because it is as if the common
market maker is aggregating scarce liquidity from each market into the
common pool (Lange and Economide, 2005; Baron and Lange, 2007).

The ability to aggregate liquidity is particularly important in introducing
a new and innovative derivatives market (Shiller, 2008). One important
reason is that there is a strong network externality effect when organizing
a financial market (Stoll, 1992). People want to trade at a place where
other people also tend to trade (Stoll, 1992; Shiller, 2008). Thus, it is
difficult for the new market to gather a sufficient number of participants
above a certain threshold to ensure smooth market operation (Shiller, 2008).
In this respect, Robert Shiller notes that pari-mutuel auctions can serve as
the springboard for new markets (Baron and Lange 2007). This approach can
help new markets aggregate sufficient liquidity to compete with previously
established markets (Baron and Lange, 2007).

\subsubsection{Price Efficiency}

Pari-mutuel mechanisms enhance price efficiency because information flows
from one market to another through the common market maker (Baron and Lange,
2007). Prices of options with the same underlying asset or variable are
closely related to one another. Hence, information in one market is relevant
to the pricing of other options. Therefore, a common market maker is more
efficient than market makers involved only in a single fragmented market.
The common market maker can use information in multiple related markets when
pricing each security.\pagebreak

\section{Images and Tables}

\begin{center}
\bigskip

\bigskip

\bigskip

{\LARGE The files for images and tables used in this paper can be found at:}

{\LARGE \ https://sites.google.com/site/heesurohacademics/marketmaking}

{\Huge \ }
\end{center}

\pagebreak

\end{document}